\documentclass[12pt]{article}
\usepackage{putex}

\usepackage{xcolor}
\colorlet{dblue}{blue!70!black}
\usepackage{url}
\usepackage[colorlinks=true, urlcolor=dblue,linktocpage=true,linkcolor=dblue,citecolor=purple]{hyperref}
\def\Z{\mathbb{Z}}
\def\ord{\mathop{\rm ord}}

\begin{document}

\preprint{BRX-TH 6306, CALT-TH 2016-042, PUPT-2516}

\title{Edge length dynamics on graphs with applications to $p$-adic AdS/CFT}
\authors{Steven S. Gubser,$^\Princeton$ Matthew Heydeman,$^\Burke$ Christian Jepsen,$^\Princeton$\\[3pt] Matilde Marcolli,$^\CaltechMath$ Sarthak Parikh,$^\Princeton$ Ingmar Saberi,$^\Heidelberg$\\[3pt] Bogdan Stoica,$^{\Brandeis,\Brown}$ and Brian Trundy$^\Princeton$\\[10pt] {\tt\normalsize ssgubser@princeton.edu, mheydema@caltech.edu, cjepsen@princeton.edu, matilde@caltech.edu, sparikh@princeton.edu, saberi@mathi.uni-heidelberg.de, bstoica@brandeis.edu, btrundy@princeton.edu}}

\institution{Princeton}{$^\Princeton$Joseph Henry Laboratories, Princeton University, Princeton, NJ 08544, USA}
\institution{Burke}{$^\Burke$Walter Burke Institute for Theoretical Physics,\cr\hskip0.06in California Institute of Technology, 452-48, Pasadena, CA 91125, USA}
\institution{CaltechMath}{$^\CaltechMath$Department of Mathematics,
\cr\hskip0.06in California Institute of Technology, 253-37, Pasadena, CA 91125, USA}
\institution{Heidelberg}{$^\Heidelberg$Mathematisches Institut,  Ruprecht-Karls-Universit\"at Heidelberg, 
\cr\hskip0.06in Im Neuenheimer Feld 205, 69120 Heidelberg, Germany}
\institution{Brandeis}{$^\Brandeis$Martin A. Fisher School of Physics, Brandeis University, Waltham, MA 02453, USA}
\institution{Brown}{$^\Brown$Department of Physics, Brown University, Providence RI 02912, USA}

\abstract{We formulate a Euclidean theory of edge length dynamics based on a notion of Ricci curvature on graphs with variable edge lengths.  In order to write an explicit form for the discrete analog of the Einstein-Hilbert action, we require that the graph should either be a tree or that all its cycles should be sufficiently long.  The infinite regular tree with all edge lengths equal is an example of a graph with constant negative curvature, providing a connection with $p$-adic AdS/CFT, where such a tree takes the place of anti-de Sitter space.  We compute simple correlators of the operator holographically dual to edge length fluctuations.  This operator has dimension equal to the dimension of the boundary, and it has some features in common with the stress tensor.}

\date{December 2016}
\maketitle

\tableofcontents

\section{Introduction}

Dynamical geometry in the bulk of anti-de Sitter space is a cornerstone of the study of the anti-de Sitter / conformal field theory correspondence (AdS/CFT).  At the linearized level, propagation of gravitons in AdS can be translated into the two-point function of the stress-energy tensor in the CFT.  At the non-linear level, dynamical geometry is involved in everything from anomalies to holographic renormalization group flows to the formation of black holes.

Recent developments \cite{Gubser:2016guj,Heydeman:2016ldy} in the study of holographic relations between field theories defined on the $p$-adic numbers and bulk dynamics defined on a regular tree graph have omitted the study of dynamical geometry in the bulk.  Different bulk {\it topologies} were considered in \cite{Heydeman:2016ldy} in connection with non-archimedean generalizations of BTZ black holes, following earlier work \cite{Manin:2002hn}; but it has generally been assumed that all edges and all vertices on the tree are locally indistinguishable.  In this paper, we want to lift this restriction by considering variable edge lengths.  More specifically, we start with an action on the tree of the form
 \eqn{GraphDynamics}{
  S_\phi = \sum_{\langle xy \rangle} {(\phi_x - \phi_y)^2 \over 2 a_{xy}^2} + 
    \sum_x V(\phi_x) \,.
 }
Here $\sum_{\langle xy \rangle}$ indicates a sum over edges (i.e.~without counting $\langle xy \rangle$ and $\langle yx \rangle$ separately), and $a_{xy}$ is the length of the edge $xy$, while $V$ is a potential for the bulk scalar field $\phi_x$.  Calculations of correlators of the operator dual to $\phi_x$ were carried out in \cite{Gubser:2016guj,Heydeman:2016ldy} with all $a_{xy}$ set equal to $1$, and these calculations have notable precursors in the literature on $p$-adic strings, for example \cite{Zabrodin:1988ep}.\footnote{Meanwhile, an apparently different approach to dynamics on the tree was advanced in \cite{Harlow:2011az}, in which a directed structure on the graph is assumed, such that each vertex has a single parent and $p$ offspring.  Then one defines a process that probabilistically assigns the state of each vertex based only on the state of its parent.  Holographic correlators can be constructed in this approach in terms of the limits of combinations of the probabilities of vertices which are many steps down along the tree.}  Now we would like to ask what interesting dynamics for the edge lengths $a_{xy}$ could be added.\footnote{Of course, one could imagine also introducing some dynamics for parameters in the potentials in $V$ that vary from vertex to vertex, but since this could be done simply by adding another field $\theta_x$ on vertices and introducing $\theta$-$\phi$ interactions, we don't think of it as such an interesting avenue.}

To get started, let's set
 \eqn{JxyDef}{
  J_e = {1 \over a_e^2} \,,
 }
where $e = xy$ is an edge.  Then $J_e$ is a ``bond strength'' or ``exchange energy'' for the edge $e$.  All our discussion focuses on Euclidean signature, in which all the bond strengths are positive.  One obvious way to make the bond strengths dynamical is to include some Gaussian white noise in the $J_e$: that is, we could draw each $J_e$ independently from a Gaussian distribution.  White noise for the $J_e$ seems quite unlike gravitational dynamics, because nearby $J_e$ don't pull on one another.  Better would be to introduce some interactions among the $J_e$ on neighboring edges by adding to the \eno{GraphDynamics} an action
 \eqn{Jdynamics}{
  S_J = \sum_{\langle ef \rangle} {1 \over 2} (J_e - J_f)^2 + \sum_e U(J_e) \,,
 }
where $\langle ef \rangle$ means a sum over neighboring edges---that is, edges which share one vertex.  If we omitted the first term in \eno{Jdynamics}, and made the potential $U$ quadratic, then the $J_e$ would be independent from one another, and we would be back to the case of Gaussian white noise (but unquenched assuming we form a partition function $Z = \int {\cal D} J {\cal D} \phi \, e^{-S_\phi - S_J}$).  In particular, we see that a quadratic term in the $U$ corresponds to a mass term for the edge variables $J_e$.  Probably for something resembling gravity, we should avoid having a quadratic term in the $U$.

While \eno{Jdynamics} is a sensible starting point, it seems {\it ad hoc}.  A key idea that will lead us to a more interesting class of edge length actions is a notion of Ricci curvature on graphs with variable edge lengths.  Closely related ideas have been developed in the mathematical literature for some time: see for example \cite{Bakry1985,OLLIVIER2009810,lin2011}.  Our main point of departure is the definition of Ricci curvature in \cite{OLLIVIER2009810,lin2011} as a function of pairs of vertices (not necessarily neighboring vertices), based on a comparison of distance between the two chosen vertices and a weighted distance between two probability distributions, each one localized near one of the chosen vertices.  Our extension of this notion of Ricci curvature to the case of variable edge lengths has some arbitrariness, so we cannot claim to have a uniquely privileged definition of the graph-theoretic Ricci curvature.  However, we do have a well motivated class of constructions with good properties, including the finding that the regular tree graph with all edge lengths equal has constant negative curvature.

The plan of the rest of this paper is as follows.  In section~\ref{PADIC} we briefly review the connection between the $p$-adic numbers and the regular tree graph with coordination number $p+1$.  Then in section~\ref{EDGE} we explain how the action \eno{Jdynamics} leads to a notion of edge Laplacian which is different from the usual one, but natural from the point of view of the so-called line graph.  Next, in section~\ref{RICCI}, we give the definition of Ricci curvature which we will use.  While our motivation is $p$-adic AdS/CFT, edge length fluctuations can be studied on more general graphs.  The particular Ricci curvature construction that we introduce depends on the graph being ``almost a tree,'' in a sense that we will make precise in section~\ref{RICCI}.  (Intuitively, what ``almost a tree'' means is that all cycles in the graph should be sufficiently long.)  We explain in section~\ref{LINEARIZED} how a linearized analysis around the regular tree reduces the Ricci curvature to the edge Laplacian of the bond strengths $J_{xy}$.  We exhibit in section~\ref{ACTION} an analog of the Einstein-Hilbert action, with a boundary term similar to the Gibbons-Hawking action.  This action leads to equations of motion which are satisfied by the regular tree with equal edge lengths, and the linearized fluctuations are controlled as expected by the edge length Laplacian.  We compute in section~\ref{CORRELATORS} the simplest holographic correlators involving edge length fluctuations.  In section~\ref{SOLUTION} we describe an exact solution to the equations of motion on a regular tree which deviates strongly from constant edge length.  We conclude in section~\ref{CONCLUSIONS} by reviewing our main results and indicating some direction for future work.  Appendix~\ref{appendix:gl2} reviews aspects of the action of the $p$-adic conformal group on the graph whose boundary is the $p$-adic numbers.  Appendix~\ref{appendix:Vladimirov} explains the Vladimirov derivative, which is a crucial construction in $p$-adic field theory and was understood in the context of bulk reconstruction \cite{Heydeman:2016ldy} to be effectively a normal derivative at the boundary of the tree.

\section{Mathematical background}
\label{BACKGROUND}

In this section we briefly review two well-known mathematical concepts.  In subsection~\ref{PADIC} we explain the Bruhat-Tits tree, a regular tree whose boundary is the $p$-adic numbers.  In subsection~\ref{EDGE} we summarize the line graph construction, which renders natural the edge Laplacian that we encounter when linearizing the graph theoretic Ricci curvature to be introduced in section~\ref{RICCI}.

\subsection{$p$-adic numbers and the Bruhat-Tits tree}
\label{PADIC}

Introductions to $p$-adic numbers requiring a minimum of technical background can be found in the recent works \cite{Gubser:2016guj,Heydeman:2016ldy} and in the earlier literature on $p$-adic string theory, notably \cite{brekke1988non}.  Here we sketch only a few of the most relevant points.

For any chosen prime integer $p$, the $p$-adic numbers $\mathbb{Q}_p$ are the completion of the rationals $\mathbb{Q}$ with respect to the $p$-adic norm, defined on $\mathbb{Q}$ so that if $a$ and $b$ are non-zero integers, neither of which is divisible by $p$, then
 \eqn{PadicNorm}{
  |x|_p = p^{-v} \qquad\hbox{when}\qquad x = p^v {a \over b} \,.
 }
By definition, $|0|_p = 0$.  We will usually drop the subscript $p$ and write $|x|$ instead of $|x|_p$ when it is obvious from context that we mean the $p$-adic norm.  The $p$-adic norm is ultrametric, meaning that $|x+y| \leq \max\{ |x|,|y| \}$.  $\mathbb{Q}_p$ is a field, with multiplication, addition, and inverses defined by continuity from their usual definitions on $\mathbb{Q}$.

Any non-zero $p$-adic number can be expressed uniquely as a series:
 \eqn{xExpress}{
  x = p^v (c_0 + c_1 p + c_2 p^2 + \ldots) \,,
 }
where $v \in \mathbb{Z}$, $c_0 \in \mathbb{F}_p^\times$, and $c_i \in \mathbb{F}_p$. Here $\mathbb{F}_p^\times$ denotes the non-zero elements in $\mathbb{F}_p$.\footnote{$p$-adic numbers in $\mathbb{Q}_p$ add and multiply \emph{with carrying}, so strictly speaking $c_0$ and $c_i$ take values in $\{ 1,\dots, p-1 \}$ and $\{ 0,\dots, p-1 \}$ respectively, and not in $\mathbb{F}_p^\times$ and $\mathbb{F}_p$. For the sake of conciseness we will suppress this technical detail in the rest of the paper.}  The infinite series in \eno{xExpress} appears to be highly divergent, but in fact it converges because the $c_i$ are bounded in $p$-adic norm, while $|p^n| = p^{-n}$.  The expansion \eno{xExpress} is reminiscent of the base $p$ representation of a real number, but it is different because it terminates to the right and may continue indefinitely to the left.

The Bruhat-Tits tree, which we denote $T_p$, can be understood informally as a graphical representation of the expansion \eno{xExpress}.  We picture an infinite regular tree with coordination number $p+1$, with a privileged path leading through it (with no back-tracking) from a boundary point that we label $\infty$ to another boundary point that we label $0$.  We describe this privileged path as the ``trunk'' of the tree.  We now consider another path (also with no back-tracking) starting from the point $\infty$ and leading to some other boundary point that we are going to associate with the $p$-adic number $x$.  This new path must run along the trunk for a while, and the location where it diverges from the trunk can be labeled by the valuation $v$ of $x$ (as it appears in \eno{xExpress}).  When we branch off the main trunk, the first step we take requires a choice out of $p-1$ possible directions, so we can label this choice by an element $c_0 \in \mathbb{F}_p^\times$.  In each subsequent step, we have to choose among $p$ possible directions, and each such choice can be labeled by an element $c_i \in \mathbb{F}_p$.  In short, we see that the data required to select the new path is in precise correspondence with the information required to specify a non-zero $p$-adic number.  Since infinite non-back-tracking paths from $\infty$ through the tree are in precise correspondence with the boundary points other than $\infty$, we can say that the boundary of the tree is $\mathbb{Q}_p \cup \{\infty\}$, which is $\mathbb{P}^1(\mathbb{Q}_p)$.\footnote{If we were attempting to be rigorous, we could have started by {\it defining} the set of boundary points as the set of semi-infinite paths (with no back-tracking) starting from some specified vertex $C$ of the tree.}

It can be shown that the Bruhat-Tits tree is a quotient space:
 \eqn{BTquotient}{
  T_p = {{\rm PGL}(2,\mathbb{Q}_p) \over {\rm PGL}(2,\mathbb{Z}_p)} \,,
 }
where $\mathbb{Z}_p$ denotes the $p$-adic integers (the completion of $\mathbb{Z}$ with respect to $|\cdot|_p$, or equivalently the set of all $x \in \mathbb{Q}_p$ with $|x|_p \leq 1$).  The quotient \eno{BTquotient} is similar to the realization of the Poincar\'e disk as ${\rm SL}(2,\mathbb{R}) / {\rm U}(1)$.  A similar construction can be given for field extensions of the $p$-adic numbers: for example, the unramified extension of degree $n$, which we denote $\mathbb{Q}_q$ with $q = p^n$, is associated with a tree $T_q = {{\rm PGL}(2,\mathbb{Q}_q) / {\rm PGL}(2,\mathbb{Z}_q)}$ with coordination number $p^n+1$.  Non-zero elements $x \in \mathbb{Q}_q$ admit an expansion of the form \eno{xExpress}, except that the finite field $\mathbb{F}_p$ is replaced by the larger finite field $\mathbb{F}_q$.  Having made such an expansion, the norm of $x$ can be defined by $|x| = p^{-v}$.

The action of ${\rm PGL}(2,\mathbb{Q}_q)$ on a number $x \in \mathbb{Q}_q$ is realized through linear fractional transformations, and in particular it includes scaling $x$ by any integer power of $p$.  Consider scaling by $p^m$ for some $m>1$.  This corresponds to an isometry of $T_q$ based on a translation along the main trunk of the tree by $m$ steps.  The group $\Gamma$ generated by this translation and its inverse is an image of $\mathbb{Z}$ inside ${\rm PGL}(2,\mathbb{Q}_q)$, and the quotient space $T_q / \Gamma  $ is analogous to the construction of the BTZ black hole as a quotient by some subgroup $\Gamma \subset {\rm SO}(3,1)$ of the three-dimensional hyperbolic plane $\mathbb{H}_3 = {\rm SO}(3,1) / {\rm SO}(3)$.  By construction, $T_q / \Gamma $ has a single cycle with $m$ links, and otherwise its structure is that of a regular tree.  It is possible to consider more complicated groups $\Gamma$, and this is precisely the direction explored in \cite{Manin:2002hn,Heydeman:2016ldy}.  It is also possible to consider more general extensions of $\mathbb{Q}_p$ than the unramified extension $\mathbb{Q}_q$, but we leave an explicit account along such lines for future work.

\subsection{An edge Laplacian}
\label{EDGE}

Consider the action \eno{Jdynamics} on a graph $G$.  For applications to $p$-adic AdS/CFT, $G$ should be the Bruhat-Tits tree $T_q$ or something close to it, but all of what we will say in this section applies to a general, connected, undirected graph $G$, provided no edge of $G$ can have both its ends on the same vertex, and between any two vertices of $G$ there is at most one edge. 

It is easy to check that the equation of motion for $J$ following from the action \eno{Jdynamics} is
 \eqn{Jeom}{
  \square J_e + U'(J_e) = 0 \,,
 }
where we define an edge Laplacian $\square$ as
 \eqn{EdgeLaplacian}{
  \square J_e \equiv \sum_{f \sim e} (J_e - J_f) \,.
 }
Here $\sum_{f \sim e}$ means the sum over all edges $f$ that share a vertex with a fixed edge $e$.  The definition \eno{EdgeLaplacian} may seem a little surprising to readers accustomed to the construction of an edge Laplacian as a square of the incidence matrix.  Let's review that construction and then see how a slight variant of it leads directly to \eno{EdgeLaplacian}.  The incidence matrix $d$ on a {\it directed} graph $G$ has rows labeled by edges and columns labeled by vertices.  It is defined so that if $e$ is an edge and $v$ is a vertex, $d_{ev} = 1$ if $e$ ends on $v$, $d_{ev} = -1$ if $e$ starts on $v$, and $d_{ev} = 0$ otherwise.  The adjoint (really just a transpose since the matrix is real) is denoted $d^\dagger$, and one can construct a natural-looking edge Laplacian on $G$ as $dd^\dagger$.  Unfortunately for us, $dd^\dagger$ depends on the choice of orientation of the edges, so it cannot be regarded as well-defined on an undirected graph $G$.  This is in contrast to the standard vertex Laplacian $d^\dagger d$, which {\it doesn't} depend on the orientation of the edges and therefore can be thought of as a natural construction on an undirected graph.

\begin{figure}[h]
  \centerline{\includegraphics[width=2.2in]{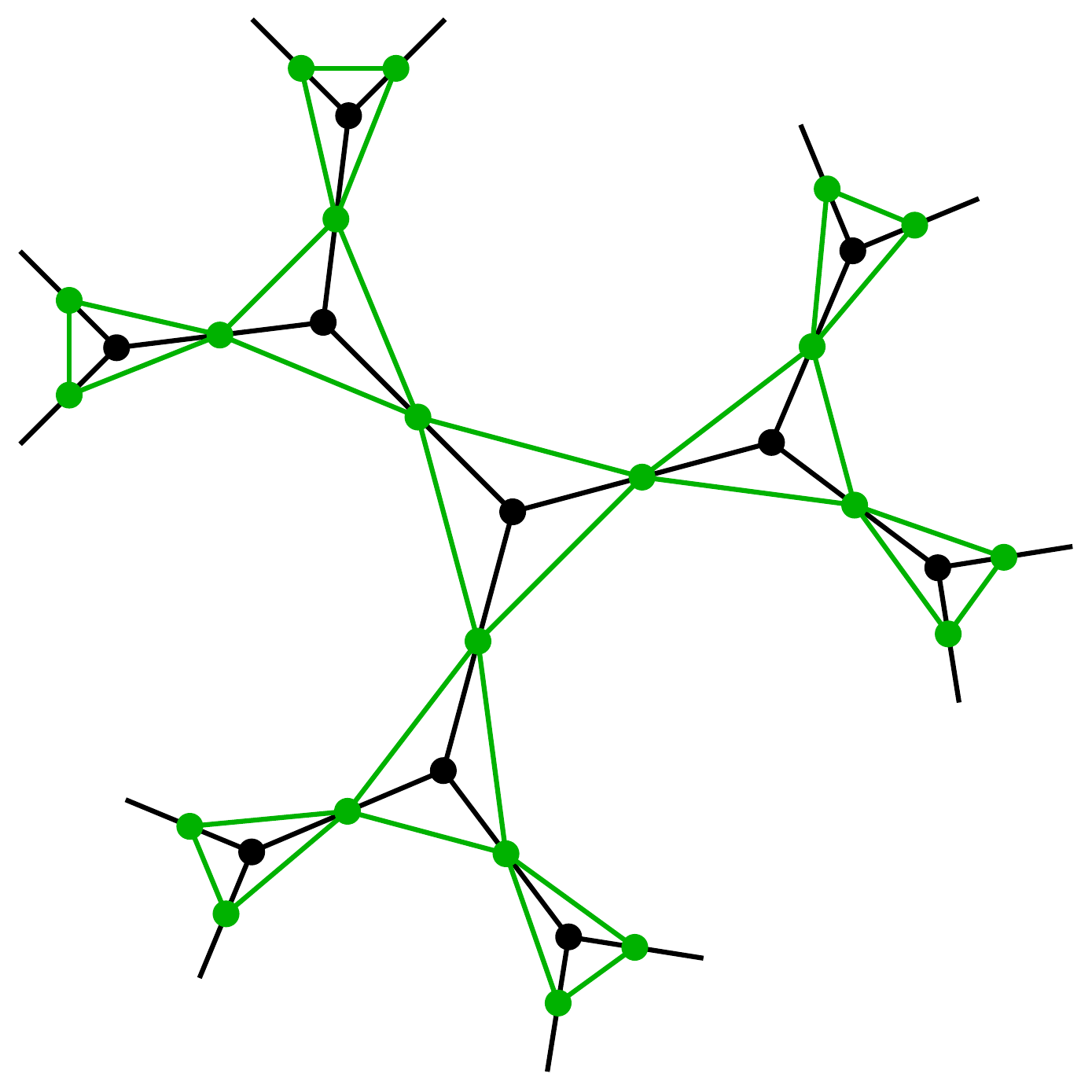}}
  \caption{A regular graph in black, and its line graph in green.}\label{LineGraph}
 \end{figure}

To make the edge Laplacian \eno{EdgeLaplacian} seem more natural, consider the line graph $L(G)$ of an undirected graph $G$.  By definition, every vertex of $L(G)$ corresponds to an edge of $G$, and two vertices of $L(G)$ are connected by an edge precisely if the corresponding two edges of $G$ meet at a vertex.  Essentially by inspection, the edge Laplacian \eno{EdgeLaplacian} on $G$ is the standard vertex Laplacian $d^\dagger d$ on $L(G)$.  It is interesting to note that the line graph of $T_q$ comprises many copies of the complete graph on $q+1$ elements, tied together by sharing each vertex between two copies: See figure~\ref{LineGraph}.

\section{Ricci curvature on graphs}
\label{RICCI}

While the action \eno{Jdynamics} seems natural enough from the point of view of dynamical models on graphs, we would prefer to have some geometrical starting point that would allow us to identify a graph-theoretic analog of the Einstein-Hilbert action.  At first it seems like a hopeless task to construct such an action on a tree graph, because the Einstein-Hilbert action involves the Ricci scalar $R$, which is usually constructed as a contraction of the Riemann tensor $R_{\mu\nu}{}^\alpha{}_\beta$.  But $R_{\mu\nu}{}^\alpha{}_\beta$ is generally thought of as the field strength of the Christoffel connection; in other words, it describes holonomies around small loops.  With no loops, it's hard to see how to define non-trivial field strengths.  To avoid this, we want to take advantage of constructions of analogs of the Ricci tensor $R_{\mu\nu}$ that do not depend on connections at all, but instead on some notion of transport distance.

\begin{figure}[h]
  \centerline{\includegraphics[width=6in]{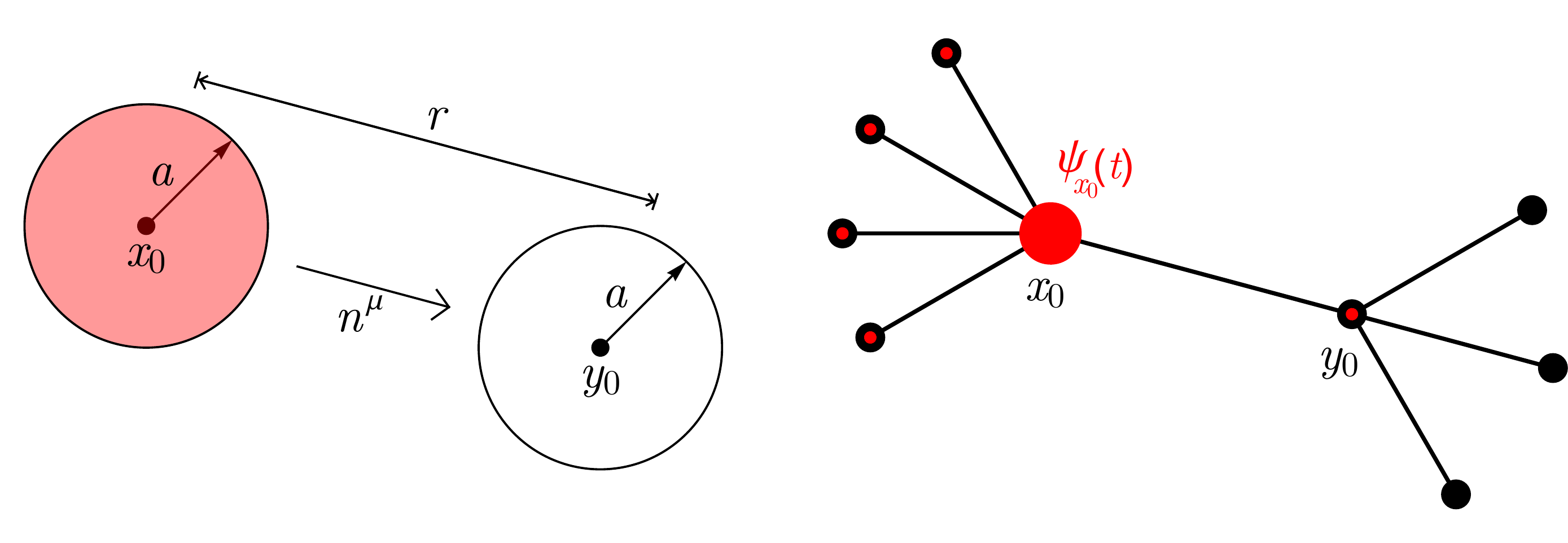}}
  \caption{Left: Small spherical neighborhoods of nearby points in a smooth manifold provide a starting point for defining Ricci curvature without first defining the Riemann tensor.  Right: A similar construction on graphs hinges on replacing the small spherical neighborhood around a point $x_0$ with a probability distribution $\psi_{x_0}(t)$ which for small $t$ is concentrated at $x_0$ with a little bit of weight on neighboring vertices.}\label{TransportRicci}
 \end{figure}

To build intuition, let's recount a standard result (see for example \cite{vonrenesse2009}) that goes in the direction we want, but which is framed in the context of a smooth $D$-dimensional manifold with a Euclidean metric which induces a distance function $d(x,y)$ between any two points on the manifold.  Given two points $x_0$ and $y_0$, separated by a small distance $r$, choose some much smaller distance $a \ll r$ and consider balls $B_{x_0}$ and $B_{y_0}$, comprising all points $x$ with $d(x,x_0) < a$ and all points $y$ with $d(y,y_0) < a$, respectively.  Let $n^\mu$ be the unit vector in the direction from $x_0$ to $y_0$; we are not concerned with exactly which tangent space $n^\mu$ lies in because we wish to use it in an asymptotic formula which can absorb $O(r)$ uncertainties in $n^\mu$.  Likewise we consider the Ricci curvature $R_{\mu\nu}$ at $x_0$ or $y_0$, or anywhere within a radius $r$ of either of these points.  There is a natural way to define a transport distance $W(B_{x_0},B_{y_0})$ between the two balls; essentially it is a weighted distance of separations of points in $B_{x_0}$ and $B_{y_0}$, but we postpone its precise definition.  Then we can form a bilocal quantity
 \eqn{kappaDefSmooth}{
  \kappa(x_0,y_0) \equiv 1 - {W(B_{x_0},B_{y_0}) \over r} = 
    {a^2 \over 2(D+2)} R_{\mu\nu} n^\mu n^\nu + O(a^3) + O(a^2 r) \,.
 }
The second equality in \eno{kappaDefSmooth} is the result we are interested in.  It tells us that the leading behavior of $\kappa(x_0,y_0)$ for small $a$ and $r$ contains all the information in $R_{\mu\nu}$---provided we are allowed to know $\kappa(x_0,y_0)$ for all possible directions of separation $n^\mu$.  See figure~\ref{TransportRicci}.

Now let's return to the definition of the transport distance $W$ appearing in \eno{kappaDefSmooth}.  Consider the so-called Wasserstein distance $W(p_1,p_2)$ between two probability measures on our smooth manifold.  We introduce the set ${\cal L}_1$ of $1$-Lipschitz functions, which are real-valued function on our smooth manifold satisfying
 \eqn{OneLipschitz}{
  |f(x) - f(y)| \leq d(x,y) \qquad\hbox{for all $x$ and $y$.}
 }
Then the Wasserstein distance is
 \eqn{WassDistance}{
  W(p_1,p_2) = \sup_{f \in {\cal L}_1} 
    \int dx \, f(x) \left[ p_1(x) - p_2(x) \right] \,.
 }
Having defined $W$ on probability measures, we define it on unit balls $B_{x_0}$ and $B_{y_0}$ by replacing each ball by the uniform probability distribution supported on the ball.  To evaluate $W(B_{x_0},B_{y_0})$ we would need $f(x)$, which to a first approximation takes the form $f(x) \approx -n_\mu x^\mu$, where $n_\mu = g_{\mu\nu} n^\nu$ and $g_{\mu\nu}$ is the Euclidean metric tensor.

When it comes to graphs, our first impulse might be to require two points $x_0$ and $y_0$ to be separated by $r \gg 1$ steps and then consider something similar to the definition \eno{kappaDefSmooth} with the balls replaced by the nearest neighbors of $x_0$ and $y_0$.  This is unattractive because our eventual aim is for $\kappa(x_0,y_0)$ to be defined for neighboring $x_0$ and $y_0$, so that $\kappa(x_0,y_0)$ can be thought of as defined on each edge; and then we hope to find in some sort of linearized analysis that $\kappa$ on edges is closely related to the edge Laplacian of fluctuations $j_{xy}$ in the bond strengths, similar to the way the Ricci tensor on a nearly flat manifold is related to the Laplacian of the metric.  So, how do we find some construction on a graph resembling a ball whose radius is much smaller than the length of a single edge?

The answer of \cite{OLLIVIER2009810,lin2011} (with closely related ideas appearing in \cite{OLLIVIER2009810}) is to consider for a fixed vertex $x_0$ a probability distribution $\psi_{x_0}(x,t)$ with most of its weight at $x=x_0$ and a small amount of weight at neighboring vertices, so that the average distance from $x_0$ of a vertex chosen from this distribution is much less than an edge length.  More precisely, for sufficiently small positive real $t$, we set
 \eqn{PsiDef}{
  \psi_{x_0}(x,t) \equiv \left\{
    \seqalign{\span\TR & \qquad\span\TT}{
     1 - {d_J(x_0) \over D_{x_0}} t & if $x=x_0$  \cr  
     {J_{x_0x} \over D_{x_0}} t & if $x \sim x_0$  \cr  
     0 & otherwise.}
    \right.
 }
We have defined
 \eqn{ddef}{
  d_J(x_0) \equiv \sum_{x \sim x_0} J_{x_0 x} \,,
 }
and, as always, we require $J_{xy} = 1/a_{xy}^2$ for all edges.  The factor of $d_J(x_0)$ in \eno{PsiDef} ensures that $\psi_{x_0}(x,t)$ is a probability distribution.  As is evident from the definition, $D_{x_0}$ is a sort of lapse function which tells us how fast the ``time'' $t$ runs at different locations on the graph.

Clearly, the definition \eno{PsiDef} is closely connected to a diffusive process.  To make this connection more precise, consider the vertex Laplacian
 \eqn{ModLap}{
  \square \phi_x \equiv \sum_{y \sim x} J_{xy} (\phi_x - \phi_y) \,.
 }
If we define a diagonal matrix on edges, $\Lambda_{ee'} = J_e \delta_{ee'}$, then it is easy to show that $\square = d^\dagger \Lambda d$,  and by inspection
 \eqn{PsiIdent}{
  \psi_{x_0}(x,t) = \left( 1 - {t \over D_{x_0}} \square_x \right) \psi_{x_0}(x,0) \,.
 }
If we want our constructions to reduce to those of \cite{lin2011} in the case when all the edge lengths $a_{xy} = 1/\sqrt{J_{xy}}$ are equal to $1$, then we should set $D_{x_0}$ to be equal to the degree of the vertex $x_0$ when all $a_{xy} = 1$.  (The degree of a vertex, usually denoted $d_{x_0}$, is the number of edges attached to it.)  An economical choice is $D_{x_0} = d_J(x_0)$, and we will make this choice in most of our subsequent development and in all our examples.  However, we cannot claim to be fixing $D_{x_0}$ from first principles.\footnote{Recent related work \cite{ChungLinYau14,ChungLinLiu2014} on Ricci curvature of weighted graphs starts with a Laplacian $\triangle = -{1 \over d_J(x)} \square$, which is suggestive of the choice $D_x = d_J(x)$.  But it is hard to make a precise comparison with our work since much of the development in \cite{ChungLinYau14,ChungLinLiu2014} follows \cite{Bakry1985} rather than \cite{OLLIVIER2009810,lin2011}; also, the focus in \cite{ChungLinYau14,ChungLinLiu2014} is on estimation of eigenvalues of $\triangle$, and the graphs of interest are usually those with non-negative Ricci curvature, whereas we are mostly interested in negative curvature.}

With the probability distributions $\psi_{x_0}(x,t)$ in place, we can follow the spirit of \eno{kappaDefSmooth} precisely.  First we define a distance function on the graph $d(x,y)$ as the minimum possible sum of edge lengths $a_e$ along a path connecting $x$ and $y$.  Then $1$-Lipschitz functions $f(x)$ defined on vertices are precisely the functions satisfying the inequality \eno{OneLipschitz}, and \eno{WassDistance} is trivially modified to
 \eqn{WassGraph}{
  W(p_1,p_2) = \sup_{f \in {\cal L}_1} \sum_x f(x) \left[ p_1(x) - p_2(x) \right] \,.
 }
Following \cite{OLLIVIER2009810,lin2011} (with variable edge lengths), we {\it define}
 \eqn{kappaDef}{
  \kappa(x,y) \equiv \lim_{t \to 0^+} \frac{1}{t} \left( 1 - 
   {W(\psi_x(t),\psi_y(t)) \over d(x,y)} \right) \,.
 }
What we mean by $\psi_x(t)$ is the probability distribution $\psi_x(t,\tilde{x})$ for all vertices $\tilde{x}$ on the graph.

\begin{figure}
  \centerline{\includegraphics[width=3in]{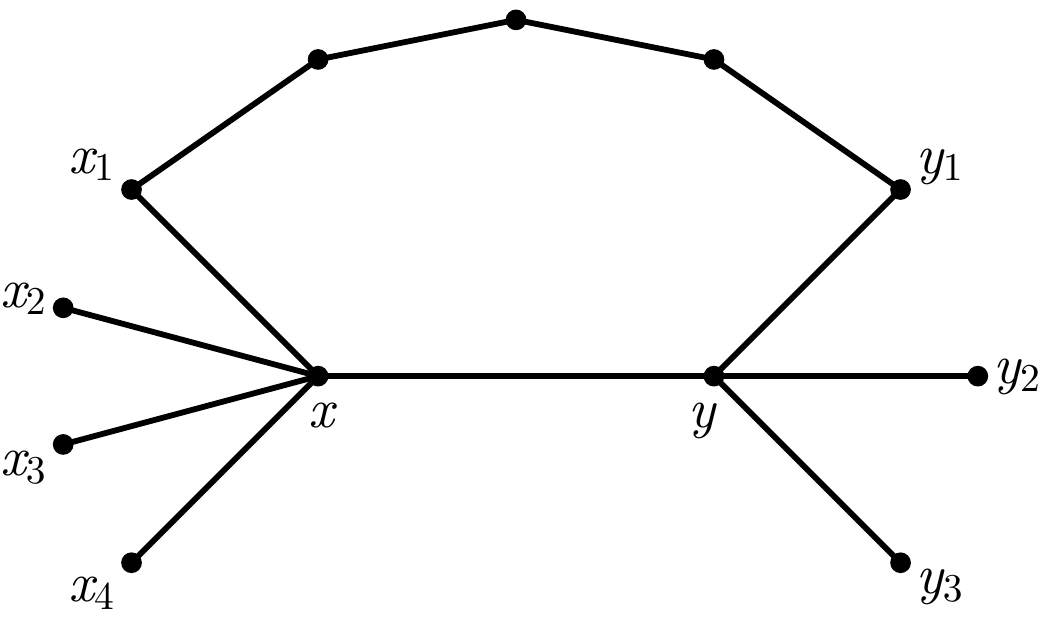}}
  \caption{Part of a graph which may qualify as ``almost a tree.''  The important criterion is that the alternate route from $x_1$ to $y_1$, passing through the top four edges, must be longer than the path from $x_1$ to $y_1$ through the edge $xy$.}\label{AlmostTree}
 \end{figure}

It is illuminating now to compute $\kappa(x,y)$ for $x$ and $y$ on opposite ends of an edge in a tree graph.  As we go through the calculation, we will see that it can be extended to graphs whose cycles are sufficiently long, in a sense that we will make precise.  We do not require for the following computation that the graph should be the Bruhat-Tits tree, but this is of course what we have in mind eventually in order to connect to $p$-adic AdS/CFT.  What makes the tree graph computation straightforward is that we can easily see what the supremizing $1$-Lipschitz function $f$ should be.  Let $x_i$ be the vertices adjacent to $x$ other than $y$, and let $y_i$ be the vertices adjacent to $y$ other than $x$.  Then we can set
 \eqn{fValues}{\seqalign{\span\TL & \span\TR &\qquad \span\TL & \span\TR}{
  f(x) &= 0 & f(y) &= -a_{xy}  \cr
  f(x_i) &= a_{xx_i} & f(y_i) &= -(a_{xy} + a_{yy_i}) \,.
 }}
An additive constant in $f$ doesn't affect the Wasserstein distance, so setting $f(x) = 0$ is just a convention.  The other choices are designed to make $f$ as positive as possible in the region where $\psi_x(t)$ has most of its weight, and as negative as possible in the region where $\psi_y(t)$ has most of its weight.  We cannot do better than \eno{fValues} because $f$ already saturates the inequality \eno{OneLipschitz} for pairs of points which are ordered in the sense of the partial ordering $x_i \preccurlyeq x \preccurlyeq y \preccurlyeq y_i$.  If our graph is not a tree, then there is the possibility that some $x_i$ might be connected to some $y_i$ by a path which is shorter (in the sense of sums of edge lengths) than the path that leads through the edge $xy$---and if that were so, then no $1$-Lipschitz function could have the values indicated in \eno{fValues}.  In order to prevent such a situation, it is sufficient to require that the graph should have no cycle with fewer than seven edges, and that the variation in edge lengths within a given cycle is by no more than a factor of $4/3$.\footnote{We could allow cycles with as few as six edges, but then no variability in edge length around the cycle can be permitted if the explicit choice \eno{fValues} for the extremizing function is to be valid.}  Then it is guaranteed that no path between an $x_i$ vertex and a $y_i$ vertex can be shorter than the one going through $xy$, and \eno{fValues} is the correct choice of a $1$-Lipschitz function that saturates the supremum in \eno{WassGraph}.  See figure~\ref{AlmostTree}.

Plugging \eno{PsiDef} and \eno{fValues} into \eno{WassGraph} and \eno{kappaDef}, we arrive at
 \eqn{RicciLocal}{
  \kappa_{xy} = {1 \over D_x a_{xy}} \left( {1 \over a_{xy}} - 
    \sum_i {1 \over a_{xx_i}} \right) + 
   {1 \over D_y a_{xy}} \left( {1 \over a_{xy}} - 
    \sum_i {1 \over a_{yy_i}} \right) \,.
 }
From now on we will refer to $\kappa_{xy}$ as given in \eno{RicciLocal} as the Ricci curvature on a graph---with the understanding that the graph is either a tree, or a graph whose loops are sufficiently large to make the calculation leading to \eno{RicciLocal} valid.  We will describe the latter sort of graph as ``almost a tree,'' keeping in mind that this apparently imprecise phrase can be rendered meaningful, for instance by imposing the previously mentioned condition that loops have to have at least seven edges, with lengths varying by no more than a factor of $4/3$.

\subsection{Negative Ricci curvature}
\label{LINEARIZED}

Consider now the Ricci curvature of the Bruhat-Tits tree with coordination number $q+1$, where $q = p^n$ and we set the length of all the edges equal to a common value $a$.  The lapse factor $D_x$ must be the same at each vertex, since in general we think of $D_x$ as a function of the edge lengths $a_{xy}$.  Let $D$ be the common value of all the $D_x$.  From \eno{RicciLocal} we have
 \eqn{RicciBT}{
  \kappa_{xy} = -{2 \over Da^2} (q-1) \,,
 }
which we understand as constant negative curvature.  If we choose $D_x = d_J(x)$, then $D = (q+1)/a^2$, and we obtain the simple result
 \eqn{RicciSpecial}{
  \kappa_{xy} = -2 {q-1 \over q+1} \,.
 }
There is a peculiar feature of \eno{RicciSpecial} which at first seems unattractive: the overall scale $a$ is undetermined.  In other words, we can scale the length of all vertices by a uniform factor, and we still have a graph with the same constant negative Ricci curvature.  We will call this feature scale freedom.  It is connected to a good feature, namely that in the linearized theory we obtain a massless equation $\square j_{xy} = 0$ for fluctuations of bond strengths around a constant $J$ solution.  Explicitly, with the choice $D_x = d_J(x)$, if we set $J_{xy} = 1 + j_{xy}$, then from \eno{RicciLocal} we find 
 \eqn{RicciLinearized}{
  \kappa_{xy} + 2 {q-1 \over q+1} = 
    -{q-3 \over 2(q+1)^2} \square j_{xy} + O(j^2) \,.
 }
Thus if we impose \eno{RicciSpecial} as an equation of motion, then at the linearized level we arrive at $\square j_{xy} = 0$, i.e.~linearized edge length fluctuations.  Admittedly, it is an odd feature that the linearized term is multiplied by a factor of $q-3$, which can be positive, negative, or even $0$ for $q$ of the form $p^n$ with $p$ prime and $n$ a positive integer.  The connection with scale freedom is that $\square j_{xy} = 0$ has as one solution $j_{xy} = {\rm constant}$, which corresponds to an infinitesimal shift in all the edge lengths.  If we broke scale freedom in a generic way, then this constant solution to the linearized equation would not exist, so the linearized equation of motion cannot be $\square j_{xy} = 0$, and edge length fluctuations would have to be massive.\footnote{A loophole in this argument is that one could perhaps arrange for the linearized equation of motion to be $\square j_{xy} = 0$, but to have terms at higher order in the fluctuations $j_{xy}$ break scale freedom.}

\subsection{A variational principle}
\label{ACTION}

While it is good to see a reasonable linearized equation of motion emerge from imposing constant negative Ricci curvature as in \eno{RicciSpecial}, we are not convinced that this is quite the optimal route to a graph theoretic version of Einstein's equations for edge length fluctuations.  The reason is that it is not clear to us how to conveniently package \eno{RicciSpecial} as the variation of an action.  Therefore, we would like to consider the action
 \eqn{RicciAction}{
  S = \sum_{\langle xy \rangle} \left( \kappa_{xy} - 2\Lambda \right) \,,
 }
which appears to be at least in the spirit of the Einstein-Hilbert action with a cosmological constant $\Lambda$.  Summing over all edges is similar to taking the trace of the Ricci tensor and then integrating over all of space.  As before, we choose $D_x = d_J(x)$, with the result that $S$ as a whole is invariant under uniformly rescaling the lengths of all edges.

The ordinary Einstein-Hilbert action is not quite a satisfactory starting point for a variational principle, because it involves second derivatives of the metric, whereas generically to get a second-order equation of motion one wants a lagrangian density which is first order in derivatives.  The well-known solution is the Gibbons-Hawking boundary term, whose effect is to cancel out the second derivative terms in the bulk Einstein-Hilbert action.  We can prescribe any (smooth) region of spacetime, add the Gibbons-Hawking term on its boundary to the Einstein-Hilbert action on its interior, and derive the Einstein equations by varying the metric inside the region while holding it fixed outside.  We would like to seek a similar augmentation of the action \eno{RicciAction}.  That is, we would like to be able to start from a large graph $G$, which is either a tree or ``almost a tree,'' isolate a subgraph $\Sigma \subset G$, and add to the action in \eno{RicciAction} a term on the boundary of $\Sigma$, after which we can vary the combined action on the interior of $\Sigma$ and recover a second order equation of motion.  Second order now means that the equation of motion should involve edges which are separated by up to two steps.  The discrete Laplace equation $\square j_{xy} = 0$ is second order because it involves $j_{xx_i}$, $j_{xy}$, and $j_{yy_i}$, and the $xx_i$ edges are two steps away from the $yy_i$ edges.

\begin{figure}[h]
  \centerline{\includegraphics[width=6in]{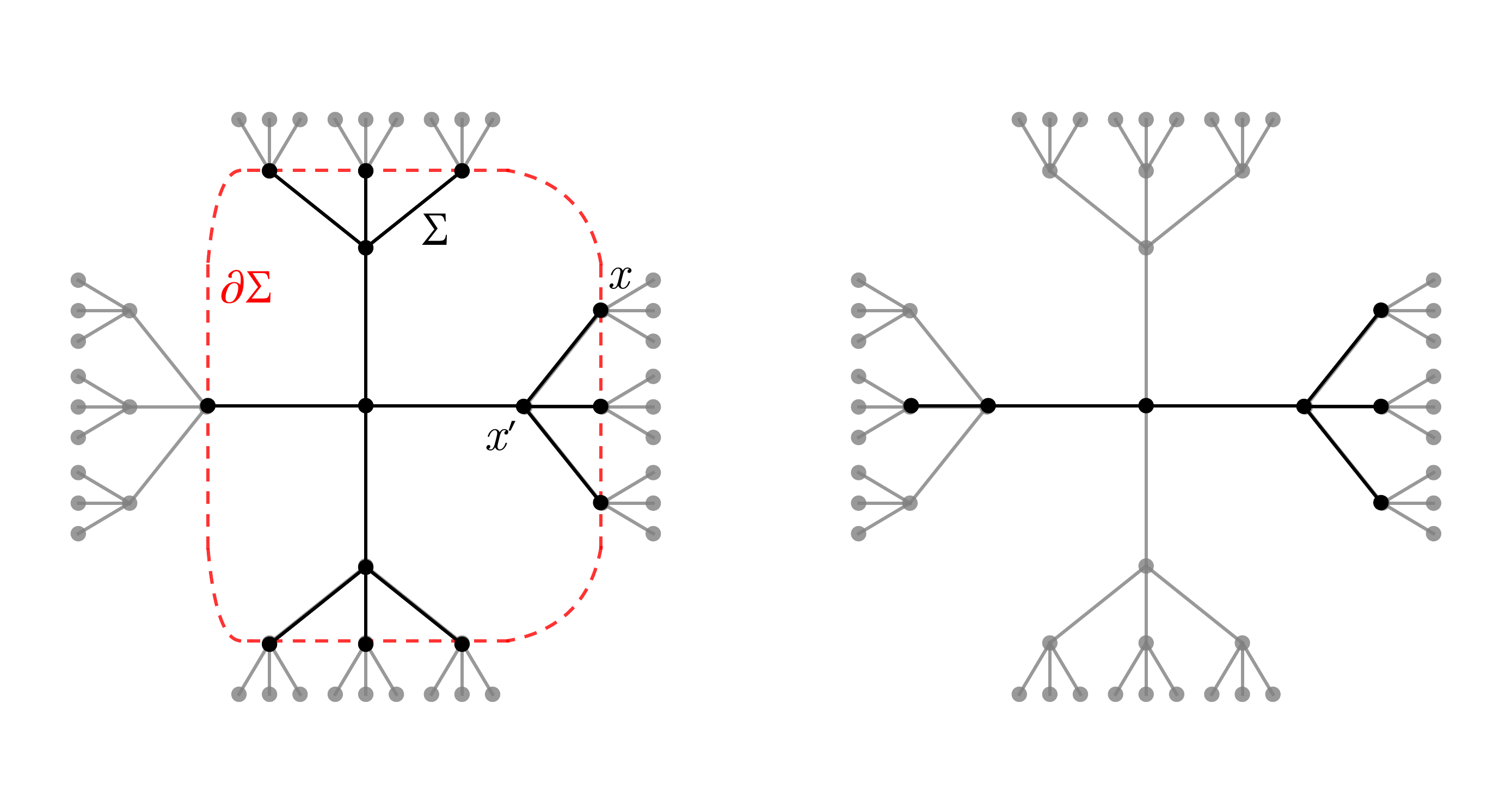}}
  \caption{Left: A fat subgraph $\Sigma$ of a regular tree.  The dashed line passes through the points on the boundary $\partial\Sigma$ of $\Sigma$.  Any point $x$ on the boundary has a unique neighbor $x'$ in the interior of $\Sigma$.  Right: A subgraph of the same regular tree which is not fat.}\label{FatSubgraph}
 \end{figure}

In order to realize the ideas of the previous paragraph concretely, we are going to put some restrictions on $\Sigma$, which we think of as a list of vertices and edges, where an edge is in $\Sigma$ iff both the vertices of that edge are in $\Sigma$.  First we require that $\Sigma$ must be a finite connected subgraph of $G$.  Consider a vertex $x \in \Sigma$ such that at least one edge connected to $x$ is {\it not} in $\Sigma$.  There must be some such vertices, because $\Sigma$ is not the whole of $G$, and we assume that $G$ is connected.  Let the collection of them be called $\partial\Sigma$.  A crucial requirement on $\Sigma$ is that for each vertex $x \in \partial\Sigma$, there is only one neighboring vertex, call it $x'$, which is in $\Sigma$, and this neighboring vertex $x'$ {\it cannot} be in $\partial\Sigma$.  We describe a subgraph $\Sigma$ that satisfies all the restrictions we have stipulated in this paragraph as a ``fat'' subgraph of $G$, and intuitively it is like a smooth finite subregion of a manifold.  Going from $x \in \partial\Sigma$ to $x'$ is like moving slightly inward from the boundary of a smooth region.  The vertices in $\Sigma - \partial\Sigma$ can be thought of as the interior of $\Sigma$.  See figure~\ref{FatSubgraph}.

It is easy to construct the subgraphs $\Sigma$ of a tree $G$ by an iterative process: starting at a vertex $x$ that is stipulated to be in the interior of $\Sigma$, we add all its neighboring vertices, and then additional vertices with the rule that once an additional vertex is included in $\Sigma$, we must either also add all its neighboring vertices not previously included in $\Sigma$ in an earlier step, or else none of them.  Of course, we must terminate this process after a finite number of steps in order to have a finite connected graph.  If $G$ has loops, then we have to be a little more careful in the choice of $\Sigma$ to make sure that $x'$ is uniquely defined for every $x \in \partial\Sigma$.  In order to be sure to have a good variational principle on all of $G$, we demand that $G$ should coincide with the union of a sequence of fat subgraphs of $G$, each of which is a subgraph of the next.

To formulate the boundary term that we need, it is convenient first to re-express \eno{RicciLocal} as
 \eqn{RicciExpress}{
  \kappa_{xy} = \kappa_{x \to y} + \kappa_{y \to x} \,,
 }
where we define a ``directed half'' of the Ricci curvature as
 \eqn{kappaDirected}{
  \kappa_{x \to y} \equiv {\sqrt{J_{xy}} \over d_J(x)} \left[ 2\sqrt{J_{xy}} - 
   c_J(x) \right] \,,
 }
and
 \eqn{CJDef}{
  c_J(x) \equiv \sum_{y \sim x} \sqrt{J_{xy}} \,.
 }
As usual we have chosen $D_x = d_J(x)$.  If $x \in \partial\Sigma$, then let's define
 \eqn{kDef}{
  k_x \equiv K_0 + \sum_{\substack{y \sim x \\ y \neq x'}} \kappa_{x \to y} \,,
 }
where $K_0$ is some constant.  Note that $d_J(x)$ and $c_J(x)$ depend on the link variables $J_{xy}$ on {\it all} the edges adjoining the vertex $x \in \partial\Sigma$, not just the edge $xx'$ belonging properly to $\Sigma$.  Likewise, $\kappa_{xx'}$ refers to all these link variables.  In formulating a boundary action in terms of $k_x$ and $\kappa_{xx'}$, we are going to regard $J_{xx'}$ as dynamical (i.e.~a quantity that we can vary), while the other $J_{xy}$---the ones just ``outside'' $\Sigma$---are known but fixed.

Now we are ready to give the action for a fat subgraph $\Sigma$ of a graph $G$ which is a tree or ``almost a tree:''
 \eqn{SwithBdy}{
  S_\Sigma = 
   \sum_{\langle xy \rangle \in \Sigma} \left( \kappa_{xy} - 2\Lambda \right) + 
    \sum_{x \in \partial\Sigma} k_x
 }
To demonstrate that this action gives rise to a well-defined equation of motion (meaning, an equation of motion which doesn't change its form on any edge when we make $\Sigma$ bigger), it is convenient first to note that we can re-express \eno{SwithBdy} as
 \eqn{Sseparated}{
  S_\Sigma = S_{\rm interior} + S_{\rm boundary}
 }
where
 \eqn{SSdefs}{
  S_{\rm interior} &\equiv \sum_{x \in \Sigma - \partial\Sigma} \sum_{y \sim x} 
    \left( \kappa_{x \to y} - \Lambda \right) 
   = \sum_{x \in \Sigma - \partial\Sigma} \sum_{y \sim x} 
    \left( {\sqrt{J_{xy}} \over d_J(x)} \left[ 
     2 \sqrt{J_{xy}} - c_J(x) \right] - \Lambda \right) \cr
  S_{\rm boundary} &= 
    \sum_{x \in \partial\Sigma} \left( -\Lambda + K_0 + 2 -
      {c_J(x)^2 \over d_J(x)} \right) \,.
 }
Varying $S_{\rm interior}$ is straightforward:
 \eqn{VaryInterior}{
  \delta S_{\rm interior} &= 
   \sum_{x \in \Sigma - \partial\Sigma} \sum_{y \sim x} \Bigg[
    {\delta J_{xy} \over 2 \sqrt{J_{xy}}} 
      {4 \sqrt{J_{xy}} - c_J(x) \over d_J(x)}  \cr 
  &\qquad\qquad\qquad{} - \sum_{z \sim x} {\delta J_{xz} \over 2 \sqrt{J_{xz}}} \left(
     {2 \sqrt{J_{xy} J_{xz}} \over d_J(x)^2} \left[ 2 \sqrt{J_{xy}} - c_J(x) \right] + 
     {\sqrt{J_{xy}} \over d_J(x)} \right) \Bigg]  \cr
  &= \sum_{x \in \Sigma - \partial\Sigma} \sum_{y \sim x} \Bigg[
    {\delta J_{xy} \over 2 \sqrt{J_{xy}}} 
      {4 \sqrt{J_{xy}} - c_J(x) \over d_J(x)}  \cr 
  &\qquad\qquad\qquad{} - \sum_{z \sim x} {\delta J_{xy} \over 2 \sqrt{J_{xy}}} \left(
     {2 \sqrt{J_{xy} J_{xz}} \over d_J(x)^2} \left[ 2 \sqrt{J_{xz}} - c_J(x) \right] + 
     {\sqrt{J_{xz}} \over d_J(x)} \right) \Bigg]  \cr
  &= \sum_{x \in \Sigma - \partial\Sigma} \sum_{y \sim x} 
    {\delta J_{xy} \over \sqrt{J_{xy}}} \left[
    \sqrt{J_{xy}} {c_J(x)^2 \over d_J(x)^2} - 
      {c_J(x) \over d_J(x)} \right]
 }
In the crucial second step of \eno{VaryInterior}, we exchange the summations over $y$ and $z$, and then relabel $y \leftrightarrow z$.  Note that $\Lambda$ does not contribute at all to the variation.  Varying $S_{\rm boundary}$ is even easier:
 \eqn{VaryBoundary}{
  \delta S_{\rm boundary} = \sum_{x \in \partial\Sigma}
    {\delta J_{xx'} \over \sqrt{J_{xx'}}} 
     \left[ \sqrt{J_{xx'}} {c_J(x)^2 \over d_J(x)^2} - 
      {c_J(x) \over d_J(x)} \right] \,.
 }
As before, the constant terms $-\Lambda$ and $K_0$ do not contribute to the variation.  Instead, the variation \eno{VaryBoundary} comes entirely from the $c_J(x)^2/d_J(x)$ term in \eno{SSdefs}, whose purpose is to produce terms in \eno{VaryBoundary} which match the form in \eno{VaryInterior}, so that in total we can write
 \eqn{VaryTotal}{
  \delta S_\Sigma = \sum_{\langle xy \rangle \in \Sigma} 
   {\delta J_{xy} \over \sqrt{J_{xy}}} \left[
    \sqrt{J_{xy}} {c_J(x)^2 \over d_J(x)^2} - {c_J(x) \over d_J(x)} + 
    \sqrt{J_{xy}} {c_J(y)^2 \over d_J(y)^2} - {c_J(y) \over d_J(y)}
    \right] \,.
 }
Thus if we define
 \eqn{gammaDef}{
  \gamma_{xy} \equiv \gamma_{x \to y} + \gamma_{y \to x}
 }
where
 \eqn{gammaDirected}{
  \gamma_{x \to y} \equiv 
    \sqrt{J_{xy}} {c_J(x)^2 \over d_J(x)^2} - {c_J(x) \over d_J(x)} \,,
 }
then the equations of motion following from the action $S_\Sigma$ are 
 \eqn{gammaZero}{
  \gamma_{xy} = 0 \,.
 }
Clearly, a regular tree, or any regular ``almost tree,'' with all $a_{xy}$ set equal to a common value $a$, gives a solution to the equations of motion \eno{gammaZero}.  If we perturb slightly around the regular tree with $a=1$ by setting $J_{xy} = 1 + j_{xy}$ for all edges, then one has immediately
 \eqn{LinearizedGamma}{
  \gamma_{xy} = {1 \over 2(q+1)} \square j_{xy} + O(j^2) \,,
 }
so that the linearized equations of motion for the edge length fluctuations are $\square j_{xy} = 0$, and this time there is no peculiar prefactor with indefinite sign like we saw in \eno{RicciLinearized}.

A feature to note is that the cosmological constant did not enter into the derivation of the equation of motion \eno{gammaZero} in any way.  This is unlike the usual Einstein-Hilbert action, where adding a cosmological constant does affect the equation of motion.  However, $\Lambda$ and $K_0$ still have a role to play in rendering the action \eno{SwithBdy} finite in the limit that we expand $\Sigma$ toward the entire graph $G$.  In order to formulate a specific prescription for obtaining a finite action, recall the way the cosmological constant enters into the usual Einstein-Hilbert plus Gibbons-Hawking action for Euclidean $AdS_3$:
 \eqn{EHArch}{
  S_\Sigma = \int_\Sigma d^3 x \, \sqrt{g} \left( R + {2 \over \ell^2} \right) - 
   2 \int_{\partial\Sigma} d^2 x \, \sqrt{-h} 
     \left( \theta + {1 \over \ell} \right) \,,
 }
where $h_{\mu\nu}$ is the induced metric on $\partial\Sigma$, and $\theta$ is the trace of the extrinsic curvature.  From \eno{EHArch} one obtains the equation of motion $R_{\mu\nu} = -{2 \over \ell^2} g_{\mu\nu}$.  Thus $R = -{6 \over \ell^2}$, and the bulk lagrangian is $R - 2\Lambda = 4\Lambda$ on shell.  To arrange an analogous situation in the action \eno{SwithBdy}, we focus on the regular tree with coordination number $q+1$ and set
 \eqn{LambdaSet}{
  \Lambda = -{1 \over 3} {q-1 \over q+1} \,,
 }
so that the ``bulk lagrangian'' $\kappa_{xy} - 2\Lambda = 4\Lambda$ when the edge length is constant.  Next we inquire what value of $K_0$ will lead to a finite limit for $S_\Sigma$ as $\Sigma$ grows.  We choose $\Sigma$ to comprise all vertices within $N$ steps of a specified vertex $C$, so that $\partial\Sigma$ is the set of vertices which are {\it exactly} $N$ steps away from $C$.  There are $n_v = (q+1) q^{N-1}$ vertices in $\partial\Sigma$, and there are
 \eqn{NeSigma}{
  N_e = (q+1) \sum_{j=0}^{N-1} q^j = {q+1 \over q-1} (q^N-1)
 }
edges in $\Sigma$ (including the ones which end on a vertex in $\partial\Sigma$).  Referring to \eno{SwithBdy}, we have
 \eqn{SwithBdyEvaluated}{
  S_\Sigma = 4\Lambda N_e + k n_v \,,
 }
where all the $k_x$ are assumed to have a common value $k$.  In order to get a finite limit for $S_\Sigma$ as $N$ becomes large, we must have
 \eqn{kFromLimit}{
  k = -4\Lambda \lim_{N \to \infty} {N_e \over n_v} = {4 \over 3} {q \over q+1} \,.
 }
Combining \eno{kDef} and \eno{kFromLimit} we find
 \eqn{FoundKzero}{
  K_0 = {q \over 3} {3q + 1 \over q+1} \,.
 }
It is easy to show that after imposing \eno{FoundKzero}, $S_\Sigma$ has a finite limit as $N \to \infty$.  The choice \eno{FoundKzero} cancels at least the leading $q^N$ divergence in a more general circumstance, where the graph $G$ under consideration is asymptotic to a regular tree with coordination number $q+1$ and constant edge length, provided we fix the cosmological constant as in \eno{LambdaSet}.

\section{Correlators}
\label{CORRELATORS}

Let's start with a total action
 \eqn{Stotal}{
  S = \sum_{\langle xy \rangle} \left( \kappa_{xy} - 2\Lambda \right) + 
   \sum_{\langle xy \rangle} {J_{xy} \over 2} (\phi_x - \phi_y)^2 + 
   \sum_x {m^2 \over 2} \phi_x^2 \,,
 }
up to boundary terms, where $\kappa_{xy}$ is defined as in \eno{RicciLocal} with our usual choice, $D_x = d_J(x)$.  From this action we would like to calculate the simplest holographic correlators of an operator ${\cal O}$ dual to $\phi$ and an operator $T$ dual to fluctuations of the bond strengths $J$.  We will focus on correlators on the Bruhat-Tits tree $T_q$, whose boundary is the unramified extension $\mathbb{Q}_q$ of $\mathbb{Q}_p$, where $q = p^n$.  Our background ``metric'' consists of setting all $J_{xy} = 1$.  We also set all $\phi_x = 0$.  The background is trivially a solution of the equations of motion following from \eno{Stotal}.  The correlators we are interested in are $\langle TT \rangle$, $\langle T{\cal O}{\cal O} \rangle$, and $\langle TTT \rangle$.  (The two-point function $\langle {\cal O} {\cal O} \rangle$ was computed already in \cite{Gubser:2016guj,Heydeman:2016ldy}.)  We will work strictly at tree level in the bulk.  We omit an overall prefactor multiplying $S$.  If such a factor were included, it would simply multiply all our correlators as a prefactor.

As a convenient parametrization, we set
 \eqn{jDef}{
  J_{xy} = 1 + j_{xy}
 }
for all edges.  We make \eno{jDef} the defining relation for $j_{xy}$, so that it is exact rather than a linearization.  To get at $\langle TT \rangle$, all we need is the part of \eno{Stotal} quadratic in the $j_{xy}$.  This quadratic action gives us propagators for $j_{xy}$, which are worked out in section~\ref{PROPAGATORS}, while $\langle TT \rangle$ itself is obtained in section~\ref{TWOPOINT}.  The three-point function $\langle T{\cal O}{\cal O} \rangle$, which we compute in \eno{MIXED}, is relatively easy because we require only the propagators for $j_{xy}$ and $\phi_x$, together with the $j_{xy} (\phi_x-\phi_y)^2$ vertex that constitutes the discrete analog of minimal coupling of the scalar to the ``metric'' represented by the bond strengths.  The three point function $\langle TTT \rangle$ is purely geometrical in the sense that only the first term in \eno{Stotal} matters.  It is a non-trivial calculation because we must expand this term to third order in the $j_{xy}$ and then track how three different types of cubic interactions among the $j_{xy}$ variables contribute to the three-point function.  Strikingly, the final result for $\langle TTT \rangle$ is zero for separated points.  We give an account of these points in section~\ref{THREEPOINT}.

\subsection{Propagators}
\label{PROPAGATORS}

We will need the distance function $d(e_1,e_2)$ between two edges on the graph $T_q$.  By definition, $d(e_1,e_2)$ is the number of vertices one must cross in order to get from $e_1$ to $e_2$.  Similarly, the distance $d(x_1,x_2)$ between two vertices on $T_q$ is the number of edges we have to cross in order to get from $x_1$ to $x_2$.  We do not account for variable edge lengths because we are perturbing around the configuration with all $J_{xy} = 1$; thus the distance function $d$ can be thought of as characterizing the background metric.  

Although our main purpose is to understand the consequences of the curvature action, we will take our calculations as far as we can with a more general action for link variables $j_e$ that includes a mass term:
 \eqn{SJagain}{
  S_J = \eta\left[ \sum_{\langle ef \rangle} {1 \over 2} (j_e - j_f)^2 +
   \sum_e {1 \over 2} m_J^2 j_e^2 \right] \,,
 }
where the prefactor $\eta$ is at this stage arbitrary.  If we expand the first term of \eno{Stotal} to quadratic order in the fluctuations $j_e$, the quadratic term agrees precisely with \eno{SJagain} provided we choose
 \eqn{etaAndM}{
  \eta = {1 \over 2(1+q)} \qquad\qquad \Delta_J = n \,.
 }
Thus we can proceed with general $\eta$ and $\Delta_J$, and at the last step specialize to massless edge length fluctuations by using \eno{etaAndM}.

Starting from the action \eno{SJagain}, we easily see that the bulk-to-bulk Green's function for fluctuations of $j_e$ should satisfy
 \eqn{GreenDefine}{
  (\square_e + m_J^2) G(e,f) = \delta_{ef} \,,
 }
where $\delta_{ef}=1$ if $e=f$ and $0$ otherwise.  One may check by direct calculation that
 \eqn{GJgeneral}{
  G_J(e,f) = -{\zeta(-\Delta_J) \zeta(2\Delta_J - 2n) \over \zeta(\Delta_J - n)}
    \hat{G}_J(e,f) \qquad\hbox{where}\qquad
    \hat{G}_J(e,f) = p^{-\Delta_J d(e,f)}
 }
solves \eno{GreenDefine}, provided $\Delta_J$ satisfies
 \eqn{DeltaJeq}{
  m_J^2 = -{1 \over \zeta(-\Delta_J) \zeta(\Delta_J-n)} \,.
 }
Here and below, we use the local zeta function
 \eqn{zetaDef}{
  \zeta(s) \equiv {1 \over 1 - p^{-s}} \,.
 }
For edge length fluctuations, where we know that $m_J=0$ from having analyzed the linearized equations of motion following from the action \eno{SwithBdy}, we set $\Delta_J=n$.  The other choice, $\Delta_J=0$, has a pathology in that the prefactor on $G_J(e,f)$ vanishes.  The correct Green's function in that case is proportional to $d(e,f)$ rather than a power of $p^{-d(e,f)}$, and this is symptomatic of logarithmic scaling behavior in the two-point function of the dual operator; compare with \cite{Zabrodin:1988ep}.

We will also need a bulk-to-boundary propagator, $K_J(e,y)$, where $y \in \mathbb{Q}_q$.  Consider the semi-infinite path $[e:y)$, where the notation $[e$ indicates that $e$ is included in the path, whereas the notation $y)$ indicates that $y$ is not.  Let $x$ be the vertex at the end of $e$ that is further from $y$, and recall from \cite{Gubser:2016guj} that we can identify $x$ as a equivalence class of points $(z,z_0)$, where $z \in \mathbb{Q}_q$ and $z_0 = p^\omega$ for some $\omega \in \mathbb{Z}$.  The equivalence relation is that we regard $(z,z_0)$ and $(z',z_0)$ as the same point iff $z' = z + z_0 n$ for some $n \in \mathbb{Z}_q$.  Then we have
 \eqn{KJexpress}{
  K_J(e,y) = p^{\Delta_J} {\zeta(\Delta_J) \zeta(2n-2\Delta_J) \over
   \zeta(2\Delta_J - n) \zeta(n-\Delta_J)} \hat{K}_J(e,y) \,,
 }
where
 \eqn{KJhat}{
  \hat{K}_J(e,y) = {|z_0|^{\Delta_J} \over |(z_0,y-z)|^{2\Delta_J}} \,.
 }
In \eno{KJhat}, $|z_0| = p^{-\omega}$ is the $p$-adic norm of $z_0$, and the norm in the denominator is $|(z_0,y-z)| \equiv \sup\{ |z_0|,|y-z| \}$.  By construction, $K_J(e,y)$ satisfies the bulk equation
 \eqn{KJsatisfies}{
  (\square_e + m_J^2) K_J(e,y) = 0 \,,
 }
and its integral over the boundary is
 \eqn{KJintegral}{
  \int_{\mathbb{Q}_q} dy \, K_J(e,y) = |z_0|^{n-\Delta_J} \,.
 }
Finally, $K_J$ satisfies the property
 \eqn{KJGJconsistency}{
  K_J(e,y) = \hat{G}_J(e,f) K_J(f,y) \,,
 }
where $f$ is any edge along the path $[e:y)$.  In section~\ref{TWOPOINT} we will need a Fourier integral of $K_J$:
 \eqn{KJFourier}{
  K_J(e,k) &\equiv \int_{\mathbb{Q}_q} dx \, \chi(ky)^* K_J(e,y)  \cr
    &= \left[ |z_0|^{n-\Delta_J} + |k|^{2\Delta_J-n} |z_0|^{\Delta_J}
      {\zeta(\Delta_J) \zeta(n-2\Delta_J) \zeta(2n-2\Delta_J) \over
       \zeta(2\Delta_J) \zeta(2\Delta_J-n) \zeta_p(n-\Delta_J)} \right]
     \gamma(kpz_0) \,,
 }
where $e$ is an edge on the path in $T_q$ from $\infty$ to $0$, and $z_0$ is the depth coordinate of the vertex of $e$ further from the boundary point $0$.  In \eno{KJFourier}, $\chi(\xi)$ is an additive character on $\mathbb{Q}_q$ with the property $\chi(\xi) = e^{2\pi i \xi}$ for rational $\xi$ (see for example \cite{Gubser:2016guj} for details on the Fourier transform over $\mathbb{Q}_q$).  The function $\gamma(\xi)$ is $1$ when $\xi \in \mathbb{Z}_q$, and $0$ otherwise.

\subsection{Two-point function}
\label{TWOPOINT}

To compute the two-point function $\langle T(z_1) T(z_2) \rangle$ for separated points $z_1,z_2 \in \mathbb{Q}_q$, we must evaluate the quadratic on-shell action \eno{SJagain} on a solution to the equations of motion.  For a solution to the equation of motion, \eno{SJagain} reduces to
 \eqn{Sonshell}{
  S_{\rm on-shell} = -{\eta \over 4} \sum_e \square j_e^2 \,.
 }
Because we are interested in separated points, we will not attempt to track boundary terms as we did for the curvature action in section~\ref{ACTION}.

We employ the familiar Fourier space method, where we label each edge $e$ by coordinates $(z_0,z)$, where $z_0 = p^\omega$ for some $\omega \in \mathbb{Z}$ and $z \in \mathbb{Q}_q$.  The meaning of this labeling is that the vertices at the ends of the edge $e$ are associated to $(z_0,z)$ and $(pz_0,z)$, where $z_0 = p^\omega$ for some $\omega \in \mathbb{Z}$, and $z \in \mathbb{Q}_q$ is defined up to replacements $z \to z+pz_0 n$ for $n \in \mathbb{Z}_q$.  Guided by \eno{KJFourier}, we set
 \eqn{FluctuationChoice}{
  j_e = \lambda_1 \chi(k_1 z) K_\epsilon(z_0,k_1) +
   \lambda_2 \chi(k_2 z) K_\epsilon(z_0,k_2) \,,
 }
where we define\footnote{A non-trivial check of \eno{FluctuationChoice} is that $\chi(kz) K_\epsilon(z_0,k)$ depends only on $e$ and not the particular $z \in \mathbb{Q}_q$ we use as the coordinate of $e$ in the boundary direction.  Only then is $j_e$ well defined as a function of the edge $e$.  To see that $\chi(kz) K_\epsilon(z_0,k)$ depends only on $e$, first note that because of the factor of $\gamma(kpz_0)$ in \eno{KepsDef}, we may assume that $|kz_0| \leq p$.  Upon replacing $z \to z + pz_0 n$ for some $n \in \mathbb{Z}_q$, the fractional part of $kz$ changes by $kpz_0 n$, and this is a $p$-adic integer since $|kpz_0 n| \leq |kpz_0| \leq 1$.  Thus $\chi(kz) \to \chi(kz + kpz_0 n) = \chi(kz)$, as desired.}
 \eqn{KepsDef}{
  K_\epsilon(z_0,k) \equiv {|z_0|^{n-\Delta_J} + 
    \zeta_J |k|^{2\Delta_J-n} |z_0|^{\Delta_J} \over
   |\epsilon|^{n-\Delta_J} + \zeta_J |k|^{2\Delta_J-n} |\epsilon|^{\Delta_J}}
   \gamma(kpz_0)
 }
and
 \eqn{zetaJDef}{
  \zeta_J = {\zeta(\Delta_J) \zeta(n-2\Delta_J) \zeta(2n-2\Delta_J) \over
       \zeta(2\Delta_J) \zeta(2\Delta_J-n) \zeta_p(n-\Delta_J)} \,.
 }
In \eno{FluctuationChoice}-\eno{KepsDef}, we have introduced a UV cutoff $\epsilon = p^\Omega$, and we prescribe a cutoff form of the on-shell action \eno{Sonshell} as follows:
 \eqn{Scutoff}{
  S_\epsilon = -{\eta \over 4} \sum_{|z_0| > |\epsilon|} \square j_e^2
    = {\eta \over 4} \left[ \sum_{|z_0| = |\epsilon|} j_e^2 - 
      \sum_{|z_0| = |\epsilon/p|} q j_e^2 \right]
 }
Each sum in square brackets is over all edges with a fixed $z_0$, as indicated.  Plugging \eno{FluctuationChoice} into \eno{Scutoff}, we obtain a regulated two-point function
 \eqn{RegTwoPoint}{
  \langle T_\epsilon(k_1) T_\epsilon(k_2) \rangle &= -{\partial^2 S_{\rm on-shell}
    \over \partial\lambda_1 \partial\lambda_2}  \cr
   &= {\eta \over 2} \left( q \sum_{|z_0|=|\epsilon/p|}
     \chi((k_1+k_2)z) \right)  \cr
   &\qquad\qquad{} \times \left[ p^{2n-2\Delta_J}
     {1 + \zeta_J p^{2\Delta_J-n} |k_1 \epsilon|^{2\Delta_J-n} \over
       1 + \zeta_J |k_1 \epsilon|^{2\Delta_J-n}}
     {1 + \zeta_J p^{2\Delta_J-n} |k_2 \epsilon|^{2\Delta_J-n} \over
       1 + \zeta_J |k_2 \epsilon|^{2\Delta_J-n}} \right]  \cr
   &\qquad{} - {\eta \over 2} \left( \sum_{|z_0|=|\epsilon|}
     \chi((k_1+k_2)z) \right)  \cr
   &= \eta \zeta_J |\epsilon|^{2\Delta_J-2n} \delta(k_1+k_2)
     {p^{2n} \over \zeta(2\Delta_J-n)} |k_1|^{2\Delta_J-n} + \hbox{(non-universal)} \,.
 }
The non-universal terms include divergent terms with no dependence on $k_1$ and $k_2$ other than $\delta(k_1+k_2)$, and also terms that are subleading relative to the term shown in the last line of \eno{RegTwoPoint} in the limit where $|k_1 \epsilon|$ and $|k_2 \epsilon|$ are small.  Referring to \cite{Gubser:2016guj}, we have
 \eqn{DivergentFourier}{
  \int_{\mathbb{Q}_q} dk \, \chi(kz) |k|^{2\Delta_J-n} = 
   {\zeta(2\Delta) \over \zeta(n-2\Delta)} {1 \over |z|^{2\Delta_J}} \,,
 }
up to divergent terms proportional to $\delta(x)$.  Thus, for separated points, we find
 \eqn{TTseparated}{
  \langle T(z_1) T(z_2) \rangle = \eta p^{2n}
   {\zeta(\Delta_J) \zeta(2n-2\Delta_J) \over \zeta(2\Delta_J-n)^2 \zeta(n-\Delta_J)}
   {1 \over |z_{12}|^{2\Delta_J}} \,,
 }
where we have attached a leg factor for the operator $T(z)$:
 \eqn{TzLeg}{
  T(z) = \lim_{\epsilon \to 0} |\epsilon|^{n-\Delta_J} T_\epsilon(z) \,.
 }
So far, in this section and in section~\ref{PROPAGATORS}, our exposition has relied  on the action \eno{SJagain}, with general $\eta$ and $\Delta_J$.  As discussed in section~\ref{PROPAGATORS}, we can specialize to the case of massless edge length fluctuations as controlled by the first term of the action \eno{Stotal} by using the values for $\eta$ and $\Delta_J$ given in \eno{etaAndM}.  Plugging these values into \eno{TzLeg} yields
 \eqn{TTfinal}{
  \langle T(z_1) T(z_2) \rangle = {p^n \over 4} {\zeta(2n) \over \zeta(n)^2}
    {1 \over |z_{12}|^{2n}} \,.
 }
Due to the factor of $|\epsilon|^{2\Delta_J-n}$ in the last line of \eno{RegTwoPoint}, there are changes of the cutoff scheme which can result in extra powers of $p^{2\Delta_J-2n}$ in the two-point function.  For instance, we could cut off the sum \eno{Scutoff} by requiring $|z_0| \geq |\epsilon|$ instead of $|z_0| > \epsilon$.  Such changes of cutoff scheme evidently do not affect \eno{TTfinal}.

\subsection{The mixed three-point function}
\label{MIXED}

To compute the mixed three-point function $\langle T(z_1) {\cal O}(z_2) {\cal O}(z_3) \rangle$ for separated points $z_1$, $z_2$, and $z_3$, we require the cubic interaction term that follows from the second term in \eno{Stotal}:
 \eqn{Sint}{
  S_{\rm int} = \sum_{\langle xy \rangle} {j_{xy} \over 2} (\phi_x - \phi_y)^2 \,.
 }
In addition to the bulk-to-boundary propagator \eno{KJexpress} for edge fluctuations, we need the bulk-to-boundary propagator for $\phi_x$, known from \cite{Gubser:2016guj}:
 \eqn{Kphi}{
  K_\phi(a,y) = {\zeta(2\Delta) \over \zeta(2\Delta-n)} \hat{K}_\phi(a,y) \,,
 }
where
 \eqn{Kphihat}{
  \hat{K}_\phi(a,y) = {|z_0|^{\Delta_\phi} \over |(z_0,y-z)|^{2\Delta_J}} \,,
 }
where now $(z_0,z)$ is understood to be a coordinate choice for the bulk vertex $a$.

The three-point function can be calculated as follows:
 \eqn{ThreePointBreakdown}{
  \langle T(z_1) {\cal O}(z_2) {\cal O}(z_3) \rangle &= 
   -\sum_{\langle ab \rangle} K_J(\langle ab \rangle,z_1) 
     \left[ K_\phi(a,z_2) - K_\phi(b,z_2) \right]
     \left[ K_\phi(a,z_3) - K_\phi(b,z_3) \right]  \cr
   &= 
   -p^{\Delta_J} {\zeta(\Delta_J) \zeta(2n-2\Delta_J) \over
     \zeta(2\Delta_J-n) \zeta(n-\Delta_J)}
     \left( {\zeta(2\Delta_\phi) \over \zeta(2\Delta_\phi-n)} \right)^2
     A_{T{\cal O}{\cal O}}(z_1,z_2,z_3)
 }
where
 \eqn{TPA}{
  A_{T{\cal O}{\cal O}}(z_1,z_2,z_3) = \hat{K}_J(\langle CC_1 \rangle,z_1)
   \hat{K}_\phi(C,z_2) \hat{K}_\phi(C,z_3) \hat{A}_{T{\cal O}{\cal O}} \,.
 }
In \eno{TPA} we have introduced the point $C$ where paths from $z_1$, $z_2$, and $z_3$ meet in $T_q$, and the adjacent point $C_1$ which is one step away from $C$ in the direction of $z_1$.  It is easy to check that
 \eqn{ThreeK}{
  \hat{K}_J(\langle CC_1 \rangle,z_1) = 
    \left| {z_{23} \over z_{12} z_{13}} \right|^{\Delta_J} \qquad
  \hat{K}_\phi(C,z_2) = \left| {z_{13} \over z_{12} z_{23}} \right|^{\Delta_\phi}
   \qquad
  \hat{K}_\phi(C,z_3) = \left| {z_{12} \over z_{13} z_{23}} \right|^{\Delta_\phi} \,,
 }
and therefore
 \eqn{KKKfactor}{
  \hat{K}_J(\langle CC_1 \rangle,z_1) \hat{K}_\phi(C,z_2) \hat{K}_\phi(C,z_3) = 
   {1 \over |z_{12}|^{\Delta_J} |z_{13}|^{\Delta_J} 
     |z_{23}|^{2\Delta_\phi-\Delta_J}} \,.
 }
The quantity $\hat{A}_{T{\cal O}{\cal O}}$ in \eno{TPA} has no dependence on the $z_i$ and comes from the summation over all edges in \eno{Sint}.  Explicit calculation of $\hat{A}_{T{\cal O}{\cal O}}$ is unilluminating, and we will quote here only the result:
 \eqn{AhatResults}{
  \hat{A}_{T{\cal O}{\cal O}} &= {\zeta(\Delta_J) \zeta(2\Delta_\phi-\Delta_J) 
     \zeta(\Delta_J+2\Delta_\phi - n) \over
    \zeta(2\Delta_J) \zeta(2\Delta_\phi)^2 \zeta(2\Delta_J-2n) \zeta(4\Delta_\phi-2n)}
      \cr &\quad{} \times
   \Big[ \zeta(\Delta_J) \zeta(\Delta_J-n) \zeta(4\Delta_\phi-2n) - 
     2 p^{\Delta_\phi-\Delta_J} \zeta(2\Delta_J) \zeta(2\Delta_J-2n)
      \zeta(2\Delta_\phi-n) \Big] \,.
 }
A significant simplification occurs when we take $\Delta_J \to n$, as appropriate for massless edge length fluctuations: then the three-point function becomes
 \eqn{FinalTOO}{
  \langle T(z_1) {\cal O}(z_2) {\cal O}(z_3) \rangle = 
    -{\zeta(n) \zeta(2\Delta_\phi) \over \zeta(2\Delta_\phi-n) \zeta(-\Delta_\phi)
      \zeta(\Delta_\phi-n)} {1 \over |z_{12}|^n |z_{13}|^n |z_{23}|^{2\Delta_\phi-n}}
      \,.
 }

\subsection{The purely geometric three-point function}
\label{THREEPOINT}

To compute the three-point function $\langle T(z_1) T(z_2) T(z_3) \rangle$ for separated points, we only need the first term in \eno{Stotal}.  Expanding this curvature action to cubic order in the fluctuations $j_{xy}$, we obtain the interaction terms
 \eqn{jjjTerm}{
  S_{\rm int} &= \sum_{\langle xy \rangle} \left[ 
   c_1 j_{xy}^3 + c_2 j_{xy}^2 \sum_{i=1}^q (j_{xx_i} + j_{yy_i}) +
   c_3 j_{xy} \sum_{1 \leq i < k \leq q}
     (j_{xx_i} j_{xx_k} + j_{yy_i} j_{yy_k}) \right]
 }
where
 \eqn{ccc}{
  c_1 = -{q (q+3) \over 4 (q+1)^2} \qquad\quad
  c_2 = {5-q \over 8 (q+1)^2} \qquad\quad
  c_3 = {1 \over 2 (q+1)^2} \,.
 }
where as usual $x_i$ denotes the vertices adjacent to $x$ other than $y$, while $y_i$ denotes the vertices adjacent to $y$ other than $x$.  Similarly to \eno{ThreePointBreakdown}-\eno{TPA}, we can easily see that
 \eqn{TTTbreak}{
  \langle T(z_1) T(z_2) T(z_3) \rangle = 
    -\left[ \prod_{i=1}^3 K_J(\langle CC_i \rangle,z_i) \right]
    \hat{A}_{TTT} \,,
 }
where $C$ is the vertex in $T_q$ where paths from $z_1$, $z_2$, and $z_3$ meet, and each $C_i$ is the vertex next to $C$ one step closer to the corresponding $z_i$.  The factor $\hat{A}_{TTT}$ has no dependence on the $z_i$, and for generic values of the coefficients $c_i$ it is non-vanishing; however, remarkably, for the particular values \eno{ccc}, we find $\hat{A}_{TTT} = 0$.  (This is for $\Delta_J=n$; in contrast to previous subsections, we do not consider general $\Delta_J$ here.)

Consider first the contribution of the $j_{xy}^3$ interaction in \eno{jjjTerm} to the three-point function: It is
 \eqn{tpfOne}{
  {\cal G}_1(z_1,z_2,z_3) = -6 \sum_e \prod_{i=1}^3 K_J(e,z_i)
   = -\left[ \prod_{i=1}^3 K_J(\langle CC_i \rangle,z_i) \right] P_1
 }
where
 \eqn{PoneExpress}{
  P_1 = 6 \sum_e \prod_{i=1}^3 h_i(e) \,.
 }
In \eno{PoneExpress} we have introduced functions
 \eqn{Gmod}{
  h_i(e) = {K_J(e,z_i) \over K_J(\langle CC_i \rangle,z_i)}
 }
on the tree.  By construction, $h_i(e)$ increases by a factor of $p^{\Delta_J}$ for each step that $e$ takes along the path from $C$ to $z_i$ in the direction of $z_i$.  But it decreases by a factor of $p^{-\Delta_J}$ for each step that $e$ takes off of this path.  Intuitively, $h_i(e)$ is like the bulk-to-bulk propagator $\hat{G}(\langle CC_i \rangle,e)$, but when the path from $\langle CC_i \rangle$ to $e$ has vertices in common with the path from $C$ to $z_i$, $h_i(e)$ includes extra positive powers of $p^{\Delta_J}$ (relative to $\hat{G}(\langle CC_i \rangle,e)$) to account for back-tracking.

Following steps similar to \eno{tpfOne} for the remaining terms in \eno{jjjTerm}, we find
 \eqn{AttoThree}{
  \hat{A}_{TTT} = \sum_{i=1}^3 c_i P_i \,,
 }
where
 \eqn{PTwoThree}{
  P_2 &= \sum_{\langle xy \rangle} \sum_{\sigma \in S_3} 
    h_{\sigma(1)}(\langle xy \rangle)
    h_{\sigma(2)}(\langle xy \rangle)
   \sum_{i=1}^q \left( h_{\sigma(3)}(\langle xx_i \rangle) + 
    h_{\sigma(3)}(\langle yy_i \rangle) \right)   \cr
  P_3 &= \sum_{\langle xy \rangle} \sum_{\sigma \in S_3} 
    h_{\sigma(1)}(\langle xy \rangle) \sum_{1 \leq i < k \leq q}
     \Big( h_{\sigma(2)}(\langle xx_i \rangle)
            h_{\sigma(3)}(\langle xx_k \rangle) 
     + h_{\sigma(2)}(\langle yy_i \rangle)
       h_{\sigma(3)}(\langle yy_k \rangle) \Big)  \,.
 }
In \eno{PTwoThree} we have summations over all permutations $\sigma$ in the symmetric group $S_3$.  The reason is that we must be able to map any permutation of the three edges $CC_i$ to the three edges involved in the interactions \eno{jjjTerm}.  A similar sum implicitly entered into \eno{PoneExpress}, but it gave only the prefactor of $6$ because the interaction term $j_{xy}^3$ doesn't distinguish among the different permutations.  The end result of performing the sums in \eno{PoneExpress} and \eno{PTwoThree} is
 \eqn{Pott}{
  P_1 = 24 p^{-2n} \zeta(n) \qquad P_2 = -48 \zeta(-n) \qquad P_3 = 24 \,,
 }
and plugging into \eno{AttoThree} results in $\hat{A}_{TTT} = 0$ upon using the coefficients \eno{ccc}.  Thus the three-point function vanishes:
 \eqn{TTTvanishes}{
  \langle T(z_1) T(z_2) T(z_3) \rangle = 0
 }
for separated points $z_1$, $z_2$, and $z_3$.  It may be noted that \eno{TTTbreak} does not account for boundary terms in the action.  Because such terms (at least, the boundary terms we found in section~\ref{ACTION}) are local on the boundary, they do not affect the result \eno{TTTvanishes} for separated points.  A proper understanding of contact terms undoubtedly does require an account of boundary terms.

\section{Solutions to the discrete Einstein equations}
\label{SOLUTION}

We saw in section~\ref{RICCI} (equations \eno{gammaDef}-\eno{gammaZero} in particular) that the discrete version of the Einstein equation takes the form $\gamma_{x \to y} + \gamma_{y \to x} = 0$, where $\gamma_{x \to y}$ is a ``directed half'' of the variation of the edge length action with respect to $J_{xy}$.  The only solutions we have exhibited so far are the trivial ones where $J_{xy}$ is constant for all edges, and these solutions trivially satisfy the stronger equations $\gamma_{x \to y} = 0$, which can be recast as
 \eqn{AsymmetricEinstein}{
  \sqrt{J_{xy}} \sum_{z \sim x} \sqrt{J_{xz}} = 
     \sum_{z \sim x} J_{xz} \,.
 }
Clearly, setting all the $J_{xy}$ to a common value solves \eno{AsymmetricEinstein} on {\it any} graph $G$, regular or not, with or without loops.  Perhaps less obviously, constant $J_{xy}$ is the only solution to \eno{AsymmetricEinstein}, provided only that $G$ is connected.  To see this, let $x$ be a fixed vertex, and sum \eno{AsymmetricEinstein} over all $y$ adjacent to $x$.  We get
 \eqn{cdDegree}{
  c_J(x)^2 = (q_x+1) d_J(x) \,,
 }
where $q_x + 1$ indicates the coordination number of the vertex $x$ (the number of edges connected to it), and $c_J(x) = \sum_{y \sim x} \sqrt{J_{xy}}$ while $d_J(x) = \sum_{y \sim x} J_{xy}$ as in previous sections.  Now define two vectors in $\mathbb{R}^{q_x+1}$:
 \eqn{bvDef}{
  \vec{v} = \left( 1,1,\ldots,1 \right) \qquad\qquad
  \vec{b} = \left( \sqrt{J_{xx_1}},\sqrt{J_{xx_2}},\ldots,
   \sqrt{J_{xx_{q_x+1}}} \right) \,.
 }
Here and below, we use $x_1,x_2,\ldots,x_{q_x+1}$ to denote the neighboring vertices of a given vertex $x$.  It is illuminating to rewrite \eno{cdDegree} as
 \eqn{CSSat}{
  (\vec{v} \cdot \vec{b})^2 = \vec{v}^2 \vec{b}^2 \,.
 }
Recalling that the Cauchy-Schwarz inequality, $(\vec{v} \cdot \vec{b})^2 \leq \vec{v}^2 \vec{b}^2$, is saturated only when $\vec{v}$ and $\vec{b}$ are linearly dependent, we see that all entries in $\vec{b}$ must in fact be identical.  Replaying the argument for each vertex $x$ in $G$, we see that the edge lengths around each vertex must be equal, and that means that $a_{xy}$ is the same for all edges in $G$ given that it is a connected graph.

We are now going to write a more explicit form of the discrete Einstein equations \eno{gammaZero} which will make it easier to find solutions with non-constant edge lengths.  In the discussion to follow, the graph $G$ can still be a general connected graph.  However, the discrete Einstein equations are well motivated (at least, according to our development) only when $G$ is ``almost a tree'' in the sense explained in section~\ref{RICCI}.  To proceed, we introduce the positive quantities
 \eqn{DefineLambda}{
  \lambda_{x \to y} \equiv {\sqrt{J_{xy}} c_J(x) \over d_J(x)} \,,
 }
and we observe that the discrete Einstein equations can be rewritten in the form
 \eqn{lambdaCircle}{
  \left(\lambda_{x\rightarrow y} - {1 \over 2}\right)^2 + 
    \left( \lambda_{y\rightarrow x} - {1 \over 2}\right)^2 = {1 \over 2}\,,
 }
whose general solution is parametrized by an angular variable $\theta_{xy} \in (-\pi/4,3\pi/4)$ (see figure~\ref{figSolutionCircle}):
\begin{figure}[t]
\centerline{
\includegraphics[width=2.5in]{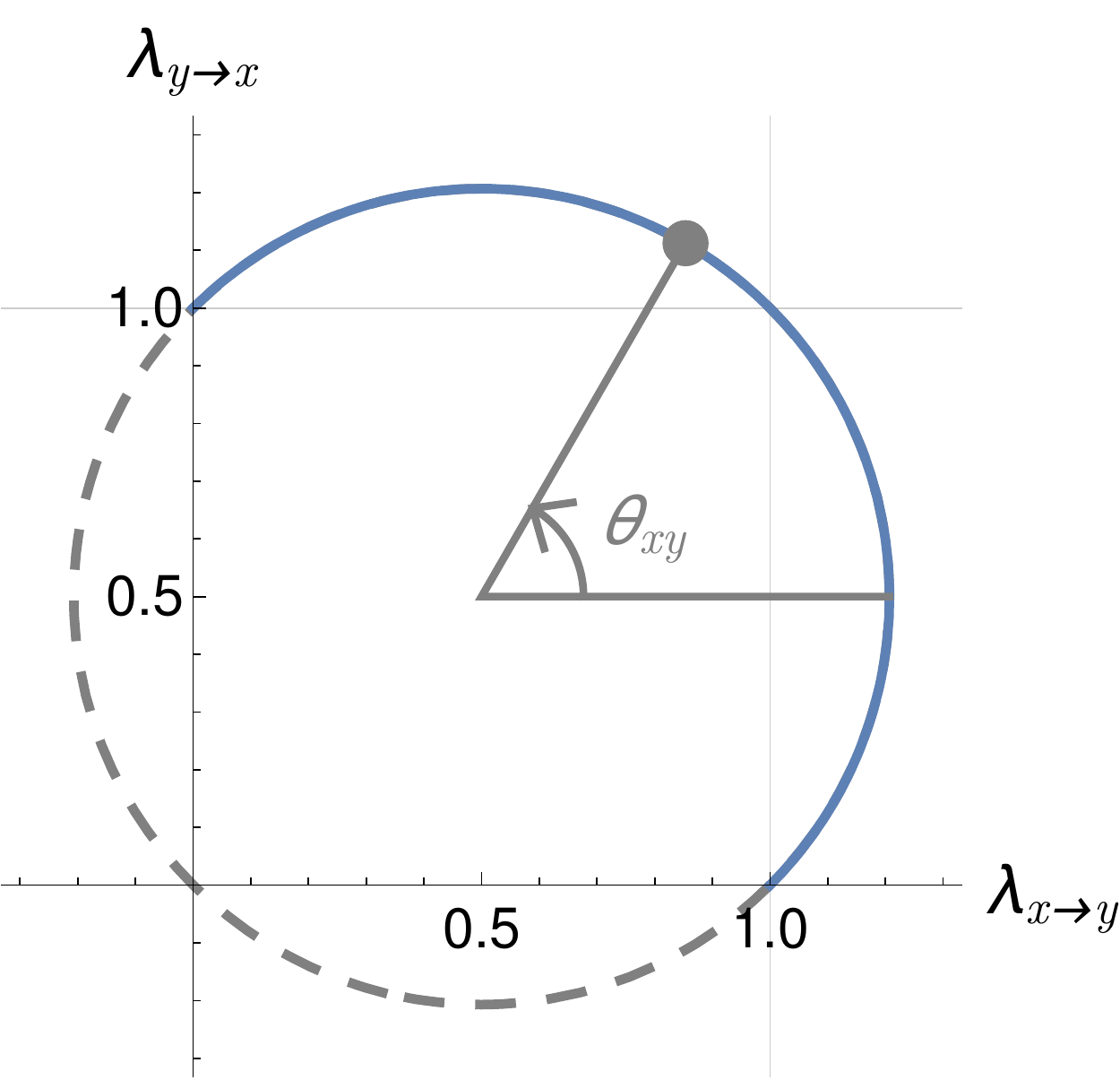}}
\caption{The ``local solution circle'' for edge $xy$. The physical solution subspace lies inside the interval $\theta_{xy} \in (-\pi/4, 3\pi/4)$ (the solid blue semi-circle).}
\label{figSolutionCircle}
\end{figure}
 \eqn{lambdaSolve}{
  \lambda_{x \to y} = {1 \over 2} + {1 \over \sqrt{2}} \cos \theta_{xy} 
    \qquad\qquad
  \lambda_{y \to x} = {1 \over 2} + {1 \over \sqrt{2}} \sin \theta_{xy} \,.
 }
The form \eno{lambdaSolve} refers implicitly to a direction on the edge $xy$, in that $\lambda_{x \to y}$ is expressed in terms of $\cos\theta_{xy}$ while $\lambda_{y \to x}$ is expressed in terms of $\sin\theta_{xy}$.  To make the notation more symmetrical, let's introduce $\theta_{x \to y} = \theta_{xy}$ and $\theta_{y \to x} = {\pi \over 2} - \theta_{xy}$.  Also introduce
 \eqn{SigmaRhoDef}{
  \sigma_{x \to y} &= \sigma(\theta_{x \to y}) \equiv 
    {1 \over 1 + \sqrt{2} \cos\theta_{x \to y}}  \cr
  \rho_{x \to y}^2 &= \rho(\theta_{x \to y})^2 \equiv 
    {q_x - \cos 2\theta_{x \to y} \over 
    (1 + \sqrt{2} \cos\theta_{x \to y})^2}
 }
for all neighboring $x$ and $y$.  Then \eno{lambdaSolve} reduces to 
 \eqn{LambdaSimpler}{
  \lambda_{x \to y} = {1 \over 2\sigma_{x \to y}} \,.
 }
Plugging \eno{DefineLambda} into \eno{LambdaSimpler} and rearranging, we wind up with
 \eqn{SphereRelation}{
  \sum_{\substack{i=1\\ i\neq k}}^{q_x+1}
    \left( \sqrt{J_{xx_i}} - \sigma_{x \to x_k} \sqrt{J_{xx_k}} \right)^2 = 
     \rho_{x \to x_k}^2 J_{xx_k} \,.
 }
(To see this, it helps to note that $\rho^2 = q\sigma^2 + 2\sigma - 1$.)

If we think of $J_{xx_k}$ as fixed, then \eno{SphereRelation} has an obvious geometrical interpretation.  Consider the space $\mathbb{R}^{q_x}$ with coordinates $(\sqrt{J_{xx_1}},\ldots,\widehat{\sqrt{J_{xx_k}}},\ldots,\sqrt{J_{xx_{q_x+1}}})$, meaning all the $\sqrt{J_{xx_i}}$ except for $\sqrt{J_{xx_k}}$.  Let $S_0$ be a sphere $S^{q_x-1}$ of radius $\rho_{x \to x_k}$ centered on the point $\sigma_{x \to x_k} (1,1,\ldots,1)$, and let $S$ be the part of $S_0$ lying in the quadrant of $\mathbb{R}^{q_x}$ where all the coordinates $\sqrt{J_{xx_i}}$ are positive.  Then \eno{SphereRelation} simply says that $S$ is the locus of possible bond strengths $J_{xx_i}$ for the edges other than $xx_k$ ending on a given vertex $x$.

To recover the constant $J_{xy}$ solutions from \eno{SphereRelation}, we set all $\theta_{x \to y} = \pi/4$, so that $\sigma_{x \to y} = 1/2$, $\rho_{x \to y}^2 = q_x/4$, and \eno{SphereRelation} is trivially satisfied for all $x$ and all neighboring $x_k$.  We would now like to exhibit a non-trivial solution on a graph with the topology of $T_q$ for odd $q$, based on the idea that half the bond strengths leading into a given vertex take one value, while the other half take a different value.  (It doesn't matter whether $q = p^n$ for some odd prime $p$.)  Pick a particular angle $\alpha \in (-\pi/4,3\pi/4)$, not equal to $\pi/4$, and set $\tilde\alpha = {\pi \over 2} - \alpha$.  Let us abbreviate notation by setting $\sigma = \sigma(\alpha)$, $\tilde\sigma = \sigma(\tilde\alpha)$, $\rho = \rho(\alpha)$, and $\tilde\rho = \rho(\tilde\alpha)$.  Then at each vertex $x$, we set
 \eqn{JEvenOdd}{
  J_{xx_i} = \left\{ \seqalign{\span\TR &\qquad\span\TT}{
   \tilde\sigma^2 J_x & for $i$ even  \cr
   \sigma^2 J_x & for $i$ odd
  } \right. \qquad\qquad
 \theta_{x \to x_i} = \left\{ \seqalign{\span\TR &\qquad\span\TT}{
   \alpha & for $i$ even  \cr
   \tilde\alpha & for $i$ odd\,,
  } \right.
 }
where the $J_x$ are as yet undetermined real positive numbers.  Already, \eno{JEvenOdd} passes a non-trivial test: namely, \eno{SphereRelation} is satisfied both for odd and even $k$, due to the  unobvious but easily verified identities
 \eqn{Unobvious}{
  {q-1 \over 2} (\tilde\sigma - \sigma \tilde\sigma)^2 + 
    {q+1 \over 2} (\sigma - \sigma \tilde\sigma)^2 &= \rho^2 \tilde\sigma^2  \cr
  {q-1 \over 2} (\sigma - \sigma \tilde\sigma)^2 + 
    {q+1 \over 2} (\tilde\sigma - \sigma \tilde\sigma)^2 &=
    \tilde\rho^2 \sigma^2 \,.
 }

What remains is to check that the vertices can be tied together so that the assignments \eno{JEvenOdd} are consistent when applied to all vertices.  Let $y$ be one of the neighbors of $x$, so that $y=x_k$ for some $k$.  It must be that $x=y_\ell$ for some $\ell$, where the $y_i$ are all the neighbors of $y$.  The edge $xx_k$ is also the edge $yy_\ell$, and we can look at consistency conditions on this edge.  The assignments of $\theta_{x \to x_i}$ in \eno{JEvenOdd} immediately lead us to conclude that $k$ and $\ell$ must have opposite parity.  This is because if $\theta_{x \to y} = \alpha$, then $\theta_{y \to x} = \tilde\alpha$ by definition of $\theta_{x \to y}$ and $\theta_{y \to x}$.  

\begin{figure}[t]
\centerline{
\includegraphics[width=3in]{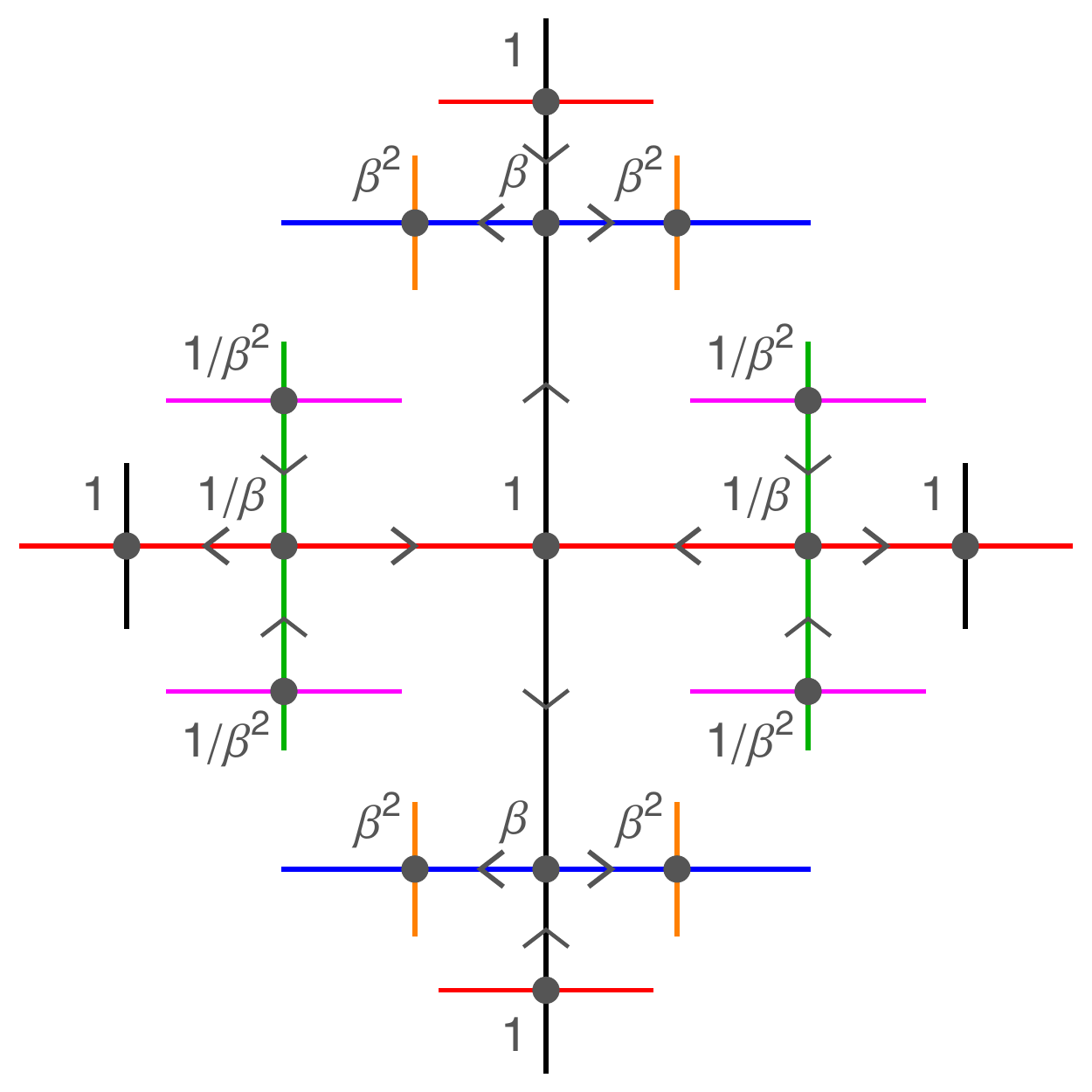}\hfill\includegraphics[width=3in]{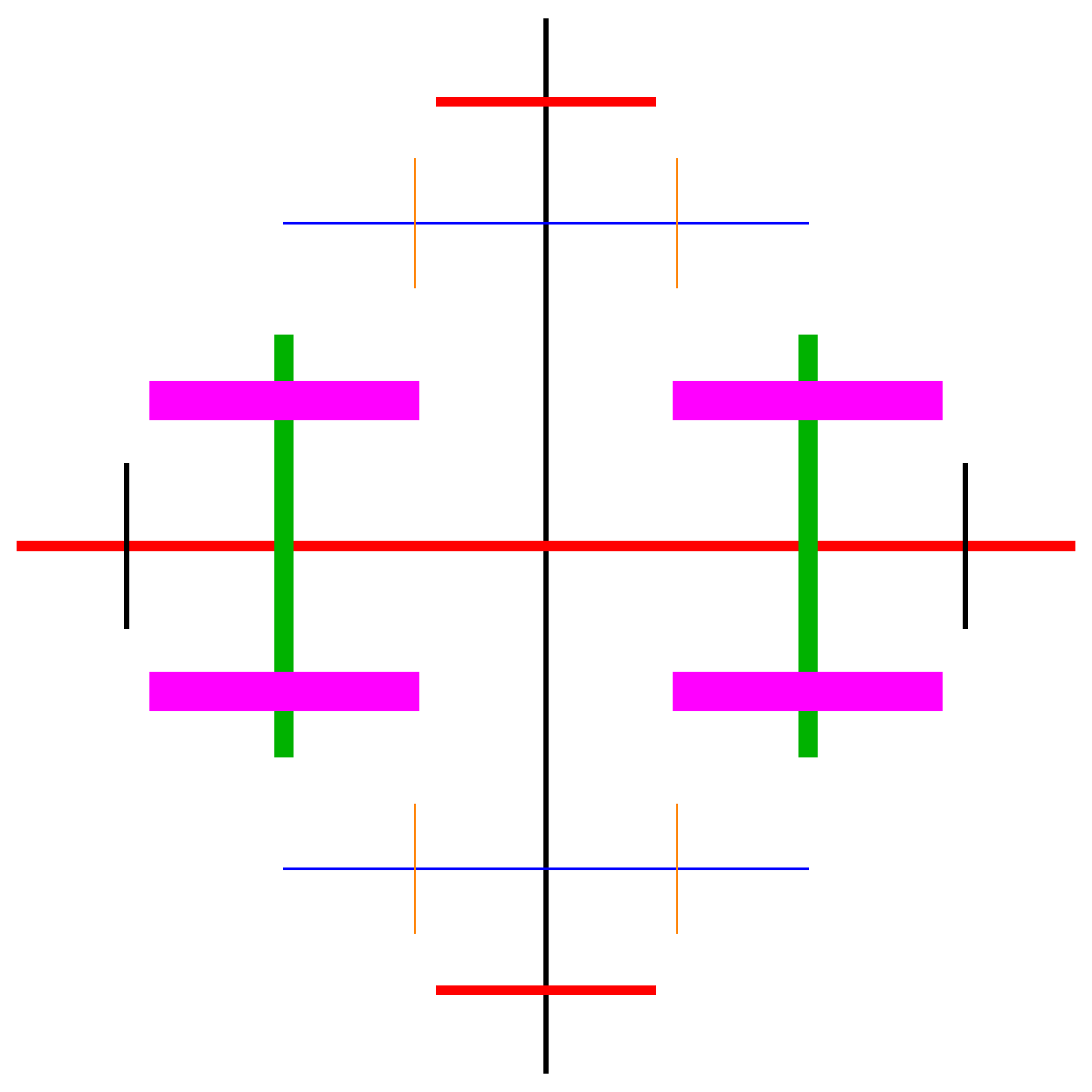}}
\caption{A regular tree (for $q = 3$) with non-constant edge lengths as described by \eno{JEvenOdd} and \eno{JzForm}. Left: Each vertex $x$ is labeled by the value $J_x$, and $\beta \equiv (\tilde\sigma/\sigma)^2$. The orientation of an edge indicates the direction in which the edge can be expressed in the form $w \rightarrow w_j$ with $j$ even. The color of the edge stands for the bond strength $J_{xy}$. Edges of the same color have equal bond strengths.  Right: The same tree, now with edges of larger width indicating larger bond strengths, taking $\beta < 1$.}\label{figNonConstantParity}
\end{figure}

Now that we have a consistent assignment of $\theta_{x \to y}$ and $\theta_{y \to x}$, we can ask about the bond strength between $x$ and $y$.  Assume $k$ is even.  Then $J_{xx_k} = \tilde\sigma^2 J_x$ from the assignments at vertex $x$, while $J_{yy_\ell} = \sigma^2 J_y$ from the assignments at vertex $y$.  But the edges $xx_k$ and $yy_\ell$ coincide: they are both the edge $xy$.  Thus we see that $J_y = (\tilde\sigma/\sigma)^2 J_x$.  If instead $k$ is odd, then the same reasoning would lead us to $J_y = (\sigma/\tilde\sigma)^2 J_x$.  Continuing, we see that if a vertex $z$ can be reached from a fixed vertex $x$ along a path where $N_{\rm even}$ of the directed links have the form $ww_i$ with $i$ even, while $N_{\rm odd}$ have the same form with $i$ odd, then
 \eqn{JzForm}{
  J_z = \left( {\tilde\sigma \over \sigma} \right)^{2(N_{\rm even} - N_{\rm odd})} 
    J_x \,.
 }
The final configuration of bond strengths is unique up to relabeling of vertices and an overall rescaling of all the $J_{xy}$.  See figure~\ref{figNonConstantParity}.  We note that the solution we have exhibited is very different from constant $J_{xy}$, in that the variation in the $J_{xy}$ is exponential with respect to the number of steps along the graph.  As a result, many paths to the boundary have finite distance, while others have an exponentially diverging distance, and still others have the linearly diverging distance that one encounters in constant $J_{xy}$ solutions.  If a distance function can be induced on the boundary through some procedure of regulation starting from distance on the graph, it would be very unlike the $p$-adic distance function $|x-y|_p$ between boundary points $x$ and $y$ in $\mathbb{Q}_p$.

Surprisingly, the non-constant edge solution just described has a constant negative Ricci curvature. Plugging in the solution given by \eno{JEvenOdd} and \eno{JzForm} in \eno{RicciLocal}  at any edge $xy$, we find
 \eqn{RicciNonConstant}{
 \kappa_{xy} = -2{q-1 \over q+1} + 1-\cos (\alpha- \pi/4)\,,
 }
 where we recognize the $q$ dependent part to be the Ricci curvature of a constant edge solution, computed in \eno{RicciSpecial}. The Ricci curvature given in \eno{RicciNonConstant} displays scale freedom just like the constant edge solution, and it is negative for all $q \geq 3$ with $-\pi/4 < \alpha < 3\pi/4$. The non-constancy of the edges simply makes the Ricci curvature less negative compared to the constant edge solution.
 
  Analogous to the construction of the BTZ black hole, we can quotient the non-uniform tree by certain abelian subgroups of the isometry group of the tree. The resulting geometry is ``almost a tree'' with precisely one cycle consisting of an even number of links. The edge lengths along the cycle are not necessarily all the same; different configurations are possible from the same non-uniform tree, depending on different choices of the abelian subgroup. We leave the detailed study of such topologies for future work.

\section{Conclusions}
\label{CONCLUSIONS}

Using the ideas of \cite{OLLIVIER2009810,lin2011}, we have formulated an action principle for edge length dynamics on a graph in terms of Ricci curvature.  The action \eno{SwithBdy} is a discrete version of the Einstein-Hilbert action with a cosmological constant and a Gibbons-Hawking boundary term, and it has a well-defined variational principle leading to discrete Einstein equations \eno{gammaZero}.  In contrast to many lattice constructions, there is no intention of taking a continuum limit, at least when we have $p$-adic AdS/CFT in view.  The Bruhat-Tits tree $T_p$, which stands in for anti-de Sitter space in $p$-adic AdS/CFT, is naturally discrete, and the obvious $p$-adic conformal symmetries act on the tree as graph isometries: see Appendix~\ref{appendix:gl2}.

While there are substantial similarities between edge length dynamics and Einstein gravity, there are some key differences.  Most notably, in our construction, we do not get spin $2$ gravitons in any obvious sense.  The field theory operator $T$ dual to edge length fluctuations on the Bruhat-Tits tree $T_p$ has a two-point function $\langle T(z) T(0) \rangle \propto 1/|z|^2$, like a scalar operator.  As discussed in \cite{Heydeman:2016ldy}, higher spin would be characterized by a more general multiplicative character.  When we generalize to the unramified extension $\mathbb{Q}_{p^n}$, which is an $n$-dimensional vector space over $\mathbb{Q}_p$, we find $\langle T(z) T(0) \rangle \propto 1/|z|^{2n}$, meaning that $T(z)$ has dimension $n$, as expected for a stress tensor; but still there is no spin.  Perhaps even more surprising, the three-point function $\langle T(z_1) T(z_2) T(z_3) \rangle$ vanishes for separated points, though this is a result which seems to depend rather sensitively on the precise construction of the Ricci curvature; in particular, it depends on our choice of the lapse factor $D_x$ to be the sum $d_J(x)$ of the bond strengths for edges adjoining the vertex $x$.

There are some good reasons for the choice $D_x = d_J(x)$.  First, it is a simple way to have our definition of Ricci curvature reduce to the one in \cite{lin2011} when all edge lengths are equal.  Second, $D_x = d_J(x)$ changes smoothly under the process of connecting or disconnecting vertices by letting the bond strength $J_{xy}$ start from or go to zero.  Third, $D_x = d_J(x)$ results in a linearized equation of motion for edge length fluctuations of the form $\square j_{xy} = 0$, whereas a more general function $D_x$ will result in a mass term for these fluctuations.  Clearly, $D_x = d_J(x)$ is the simplest analytic combination of the bond strengths with the three properties just mentioned.  Nevertheless, we should keep in mind the possibility of exploring other choices of $D_x$.

There are many directions to go from here.  The action \eno{SwithBdy} seems ideally suited for an analysis of the free energy of graphs such as the non-archimedean black holes of \cite{Manin:2002hn,Heydeman:2016ldy}.  The results of section~\ref{CORRELATORS} on correlators invite an analysis in $p$-adic field theory of what we should mean by a stress energy tensor.  While $p$-adic applications obviously privilege regular graphs with at most finitely many cycles, we can investigate a much broader class of graphs.  For example, tessellations of the Poincar\'e disk could be considered, provided all cycles are sufficiently long.  Perhaps some connection between our edge length dynamics and entanglement constructions along the lines of \cite{Heydeman:2016ldy,Cao:2016mst,Donnelly:2016qqt} could be made explicit.  We look forward to reporting on these topics in the future.

\section*{Acknowledgments}

The work of S.~Gubser, C.~Jepsen, S.~Parikh, and B.~Trundy was supported in part by the Department of Energy under Grant No.~DE-FG02-91ER40671.  The work of M.~Heydeman was supported by the Department of Energy under grant DE-SC0011632, as well as by the Walter Burke Institute for Theoretical Physics at Caltech.  M.~Marcolli is partially supported by NSF grants DMS-1201512 and PHY-1205440, and by the Perimeter Institute for Theoretical Physics. The work of B. Stoica was supported in part by the Simons Foundation, and by the U.S. Department of Energy under grant DE-SC-0009987.

\appendix

\section*{Appendices}

\section{${\rm GL}_2$ transformations of edges and vertices in a uniform tree}
\label{appendix:gl2}
Here we review some discussion of \cite{Brekke:1993gf,Casselman2014} about how subgroups of ${\rm GL}_2( \mathbb{Q}_p)$ acts on edges and vertices of the tree. For a tree of constant negative curvature with uniform edge lengths, these ${\rm GL}_2$ properties continue to hold. Much like in classifications of spin representations of the Lorentz group, we can perform a translation so that a given vertex is moved to the origin, then consider transformations that leave the origin fixed. This will tell us a bit about how fields at vertices like $\phi(x)$ and fields on edges like $J_{xy}$ behave under such transformations. 

Recall that the nodes of the Bruhat-Tits tree are lattices in $\mathbb{Q}_p^2$ modulo similarity transformations. If $u$ and $v$ form a basis of $\mathbb{Q}_p^2$, call the lattice they span $[u,v]$. If $g$ is in ${\rm GL}_2( \mathbb{Q}_p)$, acting with $g$ on the lattice takes us to another lattice $[gu, gv]$. So ${\rm GL}_2$ moves vertices around in the tree, and it turns out to also preserve edges of which there are $p+1$ per vertex. A convenient origin $x_0$ of the Bruhat-Tits tree is defined by the lattice
\begin{align}
u_0 &= (1,0) \nonumber \\
v_0 &= (0,1) \nonumber \\
x_0 &= [u_0, v_0] = \mathbb{Z}_p^2
\end{align}
The total Bruhat-Tits tree with origin $x_0$ is the coset ${\rm PGL}_2( \mathbb{Q}_p)/{\rm PGL}_2( \mathbb{Z}_p)$ (we've used $\rm P$ to take care of the similarity.) ${\rm PGL}_2( \mathbb{Z}_p)$ is the maximal compact subgroup and thus fixes the origin $x_0$. One can see that the origin is fixed by this stabilizer by explicit matrix multiplication of the basis vectors with $\mathbb{Z}_p$ coefficients; the resulting lattice will always be $\mathbb{Z}_p^2$ up to similarity. 

The nodes $1$ step from $x_0$ are labeled by elements of $\mathbb{P}^1(\mathbb{F}_p)$. This is the set of nonzero pairs $(z_1, z_2)$ in $\mathbb{Z}/p\mathbb{Z}$ modulo scalar multiplication in this group. Explicitly these are the lattices $[p u_0, v_0]$ and $[u_0 + n v_0, p v_0]$ for $n = 0, \dots, p-1$. These adjacent vertices $x \sim x_0$ are permuted by the action of ${\rm SL}(2, \mathbb{Z}_p)$. This is analogous to the ${\rm SO}(2) \subset {\rm SL}_2(\mathbb{R})$ action on the upper half plane.

Given a local field $\phi(x)$ living at a vertex in the tree, we are free to make a ${\rm GL}_2$ transformation to translate this field to the origin, $\phi(x_0)$. Further ${\rm SL}(2, \mathbb{Z}_p)$ transformations leave this invariant, and $\phi$ would appear to have the expected scalar character under the stabilizer group. For a generic field living on an edge $U_{xy}$, we can again perform a ${\rm GL}_2$ transformation to map this to $U_{x_0 x}$. As should be clear from the geometry, for a $\Lambda \in {\rm SL}(2, \mathbb{Z}_p)$, the $x$ index will transform as $U'_{x_0 x'} = \Lambda_{x x'}U_{x_0 x}$. We should not be too cavalier about calling this a spin, as in ordinary AdS different possible coordinate systems and choices of stabilizer lead to different linear combinations of AdS isometries. 

We have so far discussed the maximal compact subgroup of ${\rm GL}_2$ which fixes the origin, and we can also find a transformation which fixes a neighbor $x_1$. The neighbor is obtained by applying
\begin{equation}
\alpha = 
\begin{pmatrix}
1 & 0 \\
0 & p
\end{pmatrix}
\end{equation}
so that $\alpha(x_0) = x_1$. For $K$ a ${\rm GL}(2, \mathbb{Z}_p)$ matrix, $\alpha K \alpha^{-1}$ fixes $x_1$. We can now look for an operation which fixes both $x_0$ and $x_1$ and by construction the edge connecting them. This is found by intersection of the two stabilizer groups; $K \cap \alpha K \alpha^{-1}$ fixes the oriented edge of the tree and rotates all the branches running away from each endpoint. By conjugation every edge possess such a stabilizer. 

The fact that edge variables $U_{xy}$ transform trivially under this new set of stabilizers may make classification of spin representations more delicate. This may explain why the gravitational degrees of freedom discussed in the present work do not appear to have spin. We leave further exploration of this idea for future work.

\section{Vladimirov derivatives}
\label{appendix:Vladimirov}

In this appendix, we recall various definitions of the Vladimirov derivative operator (which is a non-local operation defined on real functions of a $p$-adic variable), and clarify some of its properties. The Vladimirov derivative is important in the context of $p$-adic AdS/CFT as the derivative operator appearing in the boundary theory, for instance in the action for the $p$-adic free boson CFT. While none of the results stated here are new, they have not (as far as we know) been clearly and explicitly summarized in previous literature.

One commonly stated definition of the Vladimirov operator $D^\alpha$ is
\begin{equation}
D^\alpha f (x) = \frac{1}{\Gamma_p(-\alpha)} \int dy\, \frac{f(y)-f(x)}{|y-x|_p^{1+\alpha}},
\label{VD-reg}
\end{equation}
where $\alpha$ is a real parameter representing the order of the derivative. This definition is puzzling for several reasons: most importantly, it's not obvious that it does what it's supposed to do (multiplication by $|k|_p$) in the Fourier domain. Furthermore, it's not clear that it has the right composition properties. We would like it to hold that
\begin{equation}
D^\alpha(D^\beta f) = D^\beta(D^\alpha f) = D^{\alpha+\beta} f.
\label{additivity}
\end{equation}
As we will show, one should understand~\eqref{VD-reg} as a regularized version of the other definition occurring in the literature: 
\begin{equation}
D^\alpha f = \pi_{-\alpha} \star f,
\label{VD-conv}
\end{equation}
where the $\star$ denotes convolution, and the family of kernels $\pi_\alpha$ are defined by 
\begin{equation}
\pi_\alpha (x) = \frac{|x|_p^{\alpha-1}}{\Gamma_p(\alpha)}.
\end{equation}
Note that plugging this definition into~\eqref{VD-conv} yields the first term, but only the first term, of~\eqref{VD-reg}. When $f(x)$ is nonzero, the second term is in fact infinite, at least for $\alpha = 1$; it diverges due to the pole in the integrand as $y\rightarrow x$. 

With regard to the additivity property~\eqref{additivity}, one would expect from the form of the definition~\eqref{VD-conv} that
\begin{equation}
\pi_\alpha \star \pi_\beta = \pi_{\alpha + \beta}.
\label{additive}
\end{equation}
In fact, this is true as long as all of the expressions involved converge; this happens when $\alpha>0$, $\beta>0$, $\alpha + \beta < 1$. The general result then follows by analytic continuation; what this amounts to is that we have to allow ourselves to resum geometric series, even if the series fail to converge. Similar behavior will occur in our analysis of the definitions of derivative. 

First of all, let's note that the class of well-behaved functions we're interested in are locally constant, and that the space of such functions is spanned by characteristic functions of $p$-adic open sets: for instance,
\begin{equation}
\gamma_\nu(x) = \begin{cases}
1, & x \in p^\nu \cdot \Z_p; \\ 
0, & x \not\in p^\nu \cdot \Z_p.
\end{cases}
\end{equation}
Since both definitions of derivative are linear and translation-invariant, we need only check their equivalence on the functions $\gamma_\nu$ to establish it in general.

Let's start with the definition by convolution, 
\begin{equation}
D \gamma_\nu (x) = \frac{1}{\Gamma_p(-1)} \int dy\, \frac{\gamma_\nu (y) }{|x-y|_p^2}.
\end{equation}
There are two cases to consider: firstly, when $|x|_p > p^{-\nu}$ (so that the pole is \textsl{outside} the support of the characteristic function and can't cause divergences), and $|x|_p \leq p^{-\nu}$. 
In the first case, $|x-y| = |x|$, and the integrand is just a constant over the region of integration; we obtain
\begin{equation}
D\gamma_\nu (x) = \frac{1}{\Gamma_p(-1)} \cdot \frac{1}{|x|_p^2} \cdot p^{-\nu} \quad (x \not\in p^\nu \cdot \Z_p),
\end{equation}
where the last factor comes from the measure of the set $p^\nu\cdot\Z_p$. 

In the second, more complicated case, there are three sub-cases to consider: $|y|$ can be strictly less than $x$, greater than, or equal. We write the integral as a sum over the circles $\ord_p y = \mu$; recall that the measure of each such circle is just $(p-1)/p^{1+\mu}$. Using the ultrametric property of the norm, and adopting the notation $\lambda = \ord_p x$, we find that
\begin{equation}
D\gamma_\nu(x) = \frac{1}{\Gamma_p(-1)} \left(
\sum_{\mu = \nu}^{\lambda-1} \frac{p-1}{p^{1+\mu}} p^{2\mu}
+ \sum_{\mu = \lambda + 1}^{\infty} \frac{p-1}{p^{1+\mu}} p^{2\lambda}
+ (\mu = \lambda \text{ term}) 
\right),
\label{sums}
\end{equation}
where we must include an extra sum over sub-circles in the $\mu=\lambda$ term, since it includes all cases $y = x + \epsilon$ where $|\epsilon| \leq |x|$. This term works out to 
\begin{equation}
\frac{p-2}{p^{1+\lambda}} \frac{1}{|x|_p^2} + \frac{p-1}{p} \sum_{\kappa > 0} \frac{1}{p^{\lambda + \kappa}} \cdot p^{2(\lambda + \kappa)} ,
\label{divpiece}
\end{equation}
and is the origin of the divergence (since $\epsilon \rightarrow 0$ is the pole $y\rightarrow x$ in the integrand). The other infinite series in~\eqref{sums} is convergent. 

To deal with this problem, we allow ourselves to resum the geometric series, even though we are obviously not within the domain of convergence: we rewrite~\eqref{divpiece} as
\begin{equation}
\frac{p-2}{p^{1-\lambda}} + \frac{p-1}{p^{1-\lambda}} \sum_{\kappa > 0} p^{\kappa}
\rightarrow 
\frac{p-2}{p^{1-\lambda}}  + \frac{p-1}{p^{1-\lambda}}\frac{p}{1-p}
= -\frac{2}{p} \cdot p^\lambda
 ,
\end{equation}
leading to the final result
\begin{equation}
D\gamma_\nu(x) = - \frac{p^{\nu - 1}}{\Gamma_p(-1)} \quad (x \in p^\nu \cdot \Z_p).
\end{equation}
The regularization we performed amounts to subtracting the infinite constant $\sum_{\kappa\in \Z} p^\kappa$, since
\begin{equation}
- \sum_{\kappa\leq 0} p^\kappa = - \frac{1}{1-p^{-1}} = \frac{p}{1-p}.
\end{equation}
A moment's thought shows that this infinite sum is just the term 
\begin{equation}
\int dy\, \frac{\gamma_\nu(x)}{|y-x|_p^2}
\end{equation}
that appears in the alternative definition~\eqref{VD-reg}. The reader can easily check that repeating the calculation using the definition~\eqref{VD-reg} yields exactly the same answer, but all quantities that appear are finite and no further regularization is required.

Now, in order to check that the regularized Vladimirov derivative~\eqref{VD-reg} satisfies the desired additivity property~\eqref{additivity}, one can simply check for an arbitrary characteristic function that 
\begin{equation}
D^\alpha (D^\beta \gamma_\nu) = D^{\alpha + \beta} \gamma_\nu,
\end{equation}
when the regularized definition~\eqref{VD-reg} is used. The general result will then follow by translation invariance and linearity. First one must generalize the above calculation to general values of the parameter $\alpha$. This is straightforward to do, and the result is
\begin{equation}
D^\alpha \gamma_\nu (x) = 
\begin{cases}
+ \frac{1}{\Gamma_p(-\alpha)}  \cdot \frac{p^{-\nu}}{|x|_p^{1+\alpha}}, & x \not\in p^\nu \cdot \Z_p, \\
- \frac{1}{\Gamma_p(-\alpha)}  \cdot  \frac{p-1}{p} \cdot \frac{p^{\alpha\nu}}{p^\alpha - 1} & x \in p^\nu \cdot \Z_p.
\label{Dalpha}
\end{cases}
\end{equation} 
To make this a bit more transparent, we still obtain a constant when $x$ is inside the support of~$\gamma_\nu$, and a decaying function (with opposite sign) when $x$ is outside. However, the value of the constant is a function of both of the parameters $\nu$ and~$\alpha$.

We then must take a further Vladimirov derivative of~\eqref{Dalpha}. The first case to consider is when the point $x$ lies inside $p^\nu\cdot\gamma_\nu$; as a reminder, we expect to get a constant with no $x$ dependence in this case. The integrand is then only nonzero when $y$ lies outside of that region, and we can evaluate the integral as
\begin{align}
D^\beta(D^\alpha \gamma_\nu)(x) &= \frac{1}{\Gamma_p(-\beta)} \int dy\, \frac{D^\alpha \gamma_\nu (y) - D^\alpha \gamma_\nu (x) }{|y-x|_p^{1+\beta}} \\
&= \frac{1}{\Gamma_p(-\beta) \Gamma_p(-\alpha)} \sum_{\mu<\nu} 
\frac{p-1}{p} p^{\beta\mu} \left[ p^{-\nu} p^{\mu(1+\alpha)} + \frac{p-1}{p} \frac{p^{\alpha \nu}}{p^\alpha - 1} \right]. 
\end{align}
Removing all non-$\mu$-dependent terms, the sum in the second term is just $\sum_{\mu<\nu} p^{\beta \mu}$, which evaluates to $p^{\beta\nu}/(p^\beta - 1)$. The sum in the first term is the same, except that $\beta$ is replaced by $(1+\alpha + \beta)$. Putting it all together, the result is
\begin{align}
D^\beta(D^\alpha \gamma_\nu)(x) &= \frac{1}{\Gamma_p(-\beta) \Gamma_p(-\alpha)} 
\left[
\frac{p-1}{p} \frac{p^{(\alpha+\beta)\nu}}{p^{1+\alpha+\beta} - 1}
+ \left( \frac{p-1}{p}\right) ^2 \frac{p^{\alpha \nu}}{p^\alpha - 1} \frac{p^{\beta \nu}}{p^\beta - 1} 
\right] \\
&= \frac{p^{(\alpha+\beta)\nu}}{\Gamma_p(-\beta) \Gamma_p(-\alpha)} \frac{p-1}{p}
\left[ \frac{1}{p^{1+\alpha+\beta} - 1} + \frac{p-1}{p} \frac{1}{(p^\alpha - 1)(p^\beta - 1)} \right].
\end{align}
A somewhat tedious computation (which is most easily done using Mathematica) shows that the coefficient reduces to 
the expected form:
\begin{equation}
D^\beta(D^\alpha \gamma_\nu)(x) = - \frac{1}{\Gamma_p(-\alpha-\beta)} \frac{p-1}{p} \frac{p^{(\alpha+\beta)\nu}}{p^{\alpha+\beta}-1}.
\end{equation}

In treating the second case, we'll use the same notation we have throughout; in particular, $\lambda = \ord_p x$. In this case, the domain of integration is not restricted, and the integrand is nonzero everywhere except on the circle $|y|=|x|$. There are three qualitatively different regions in the sum (as concerns the behavior of the integrand): where $\mu \in (-\infty, \lambda)$, $(\lambda,\nu)$, and $[\nu, \infty)$, respectively. Splitting these up and denoting them by $A$, $B$, and~$C$, we have
\begin{equation}
D^\beta(D^\alpha \gamma_\nu)(x) = \frac{1}{\Gamma_p(-\beta) \Gamma_p(-\alpha)} \frac{p-1}{p}
\left[ A + B + C
\right],
\label{splitting}
\end{equation}
where the individual sums are as follows: Firstly,
\begin{equation}
A = p^{-\nu} \sum_{\mu<\lambda} p^{\beta\mu} \left( p^{\mu(1+\alpha)} - p^{\lambda(1+\alpha)} \right)
= p^{-\nu} p^{(1+\alpha+\beta)\lambda} \left( \frac{p^{(1+\alpha+\beta)}}{p^{1+\alpha+\beta} - 1} 
- \frac{p^{\beta}}{p^{\beta} - 1}\right).
\end{equation}
(We have actually performed the sum for $\mu\leq \lambda$; it makes no difference, since the summand vanishes at $\mu=\lambda$, but allows us to write the result in a more convenient form.) Next, we can evaluate
\begin{equation}
B = p^{-\nu} p^{\lambda(1+\beta)} \sum_{\mu = \lambda + 1}^{\nu - 1} 
p^{-\mu} \left(  p^{\mu(1+\alpha)} - p^{\lambda(1+\alpha)} \right)
,
\end{equation}
which amounts to
\begin{equation}
B = p^{-\nu} p^{\lambda (1+\beta)} \left( 
\frac{p^{\alpha\nu}- p^{\alpha(\lambda + 1)}}{p^\alpha - 1} 
- p^{\lambda(1+\alpha)} \frac{p^{-\lambda} - p^{1-\nu}}{p-1}
\right)
\end{equation}
Last of all, we look at the region where $y$ lies inside the support of~$\gamma_\nu$:
\begin{align}
C &= p^{\lambda(1+\beta)} \left( - \frac{p-1}{p} \frac{p^{\alpha\nu}}{p^\alpha - 1} - p^{-\nu} p^{\lambda(1+\alpha)} \right) \sum_{\mu = \nu}^\infty p^{-\mu}  \\
&= p^{\lambda(1+\beta)} \left( - \frac{p-1}{p} \frac{p^{\alpha\nu}}{p^\alpha - 1} - p^{-\nu} p^{\lambda(1+\alpha)} \right) \frac{p^{-\nu} \cdot p}{p-1} \\
&= - \frac{p^{(\alpha-1)\nu} }{p^\alpha - 1} p^{\lambda (1+\beta)}
- \frac{p}{p-1}p^{\lambda(2+\alpha+\beta)} p^{-2\nu}.
\end{align}
Looking closely, we see that these two terms precisely cancel with two of the four terms appearing in~$B$ (the first and the last, after expanding the numerators)! This simplifies things greatly, as we can write 
\begin{equation}
B+C = - p^{-\nu} p^{\lambda ( 1 + \beta + \alpha )}  
\left( \frac{p^\alpha}{p^\alpha - 1} + \frac{1}{p-1} \right) ,
\end{equation}
and finally, gathering all terms together,
\begin{equation}
A+B+C = - p^{-\nu} p^{\lambda ( 1 + \beta + \alpha )}  
\left( \frac{p^\alpha}{p^\alpha - 1} 
+ \frac{p^{\beta}}{p^{\beta} - 1}+ \frac{1}{p-1} - \frac{p^{(1+\alpha+\beta)}}{p^{1+\alpha+\beta} - 1}  \right) .
\label{ABC}
\end{equation}
The form of the overall coefficient is familiar by now, as it has come up in several of these verifications. Plugging~\eqref{ABC} back into~\eqref{splitting} and simplifying the coefficient, we obtain the expected result:
\begin{align}
D^\beta(D^\alpha \gamma_\nu)(x) &= -  \frac{p^{-\nu} p^{\lambda (1+\alpha+\beta)} }{\Gamma_p(-\beta) \Gamma_p(-\alpha)} \frac{p-1}{p} \left( \frac{p^\alpha}{p^\alpha - 1} 
+ \frac{p^{\beta}}{p^{\beta} - 1}+ \frac{1}{p-1} - \frac{p^{(1+\alpha+\beta)}}{p^{1+\alpha+\beta} - 1}  \right) \\
&= + \frac{1}{\Gamma_p(-\alpha-\beta)}\frac{p^{-\nu}}{|x|_p^{1+\alpha+\beta}}.
\end{align}
It follows that the regularized Vladimirov derivative obeys the additivity property~\eqref{additivity} on the nose. This is not at all apparent from the form of the definition! One could have imagined that the various infinite terms that are subtracted to regularize the convolutions~\eqref{VD-conv} and~\eqref{additive} fail to cancel out, and spoil the composition law. Miraculously, this does not happen, and the regularized Vladimirov operator behaves as one would like.

\bibliographystyle{ssg}
\bibliography{trees}
\end{document}